\RequirePackage{fix-cm}
\documentclass[twocolumn]{svjour3}          
\smartqed  
\usepackage{graphicx}
\usepackage{subfigure}
\usepackage{epsfig}
\usepackage{hyperref}
\usepackage{framed}
\usepackage{psfrag}
\usepackage{tikz} 
\usepackage[symbol*]{footmisc}
\usepackage{color}
\usepackage{amssymb}
\usepackage{amsmath}
\usepackage{url}
\usepackage{listings}
\usepackage{algorithmic}
\usepackage{units}
\usepackage{float}
\journalname{Brazilian Journal of Physics}
\begin{document}

\sloppy

\title{Simulating the interaction of a non-magnetized planet with the stellar wind produced by a sun-like star using the FLASH Code}

\titlerunning{ }

\author{Edgard~F.~D.~Evangelista \and
        Oswaldo~D.~Miranda \and
        Odim~Mendes \and
        Margarete~O.~Domingues}

\authorrunning{ }

\institute{Edgard F. D. Evangelista \at
              Instituto Nacional de Pesquisas Espaciais (INPE) -- Divis\~{a}o de Astrof\'{i}sica, Av. dos Astronautas 1758, S\~{a}o Jos\'{e} dos Campos, 12227-010 SP, Brazil  \\
              \email{edgard.freitas.diniz@gmail.com (corresponding author)}
           \and
            Oswaldo~D.~Miranda \at
              INPE -- Divis\~{a}o de Astrof\'{i}sica \\
              \email{oswaldo.miranda@inpe.br}
           \and
            Odim~Mendes \at
              INPE -- Divis\~{a}o de Geof\'{i}sica Espacial\\
              \email{odim.mendes@inpe.br}
           \and
            Margarete~O.~Domingues \at
              INPE -- Laborat\'{o}rios Associados de Computa\c{c}\~{a}o e Matem\'{a}tica Aplicada \\
              \email{margarete.domingues@inpe.br}
}

\date{Received: date / Accepted: date}

\maketitle

\begin{abstract}
The study of the interaction between solid objects and magnetohydrodynamic (MHD) fluids is of great importance in physics as consequence of the significant phenomena generated, such as planets interacting with stellar wind produced by their host stars.
There are several computational tools created to simulate hydrodynamic and MHD fluids, such as the FLASH code. 
In this code there is a feature which permits the placement of rigid bodies in the domain to be simulated. 
However, it is available and tested for pure hydrodynamic cases only. Our aim here is to adapt the existing resources of FLASH to enable the placement of a rigid body in MHD scenarios and, with such a scheme, to produce the simulation of a non-magnetized planet interacting with the stellar wind produced by a sun-like star. 
Besides, we consider that the planet has no significant atmosphere. We focus our analysis on the patterns of the density, magnetic field and velocity around the planet, as well as the influence of the viscosity on such patterns. At last, an improved methodological approach is available to other interested users.

\end{abstract}

\section{Introduction}

The simulation of rigid bodies interacting with fluids is a problem of great interest in physics as consequence of the significant phenomena generated.
As examples of an application of such a problem, one may cite aerodynamic studies of mechanical structures such as airfoils and planets interacting with stellar winds. 
In the literature, one may find examples of approaches to the problem in question: in \cite{takahashi:2002}, the authors describe the modeling of the interaction of a fluid with a rigid body, where they use the Cubic Interpolated Propagation (CIP) to simulate the fluid itself and the Volume of Solid (VoS) to handle the interaction of the body with the fluid; in \cite{takashi:1992}, it is shown a computational approach to solve problems of rigid objects in contact with viscous incompressible fluids, in which the authors used the arbitrary Lagrangian-Eulerian method and the streamline-upwind/Petrov-Garlerkin finite element volume scheme.

It is worth bearing in mind that the examples such as the ones discussed above involved pure hydrodynamic scenarios only. 
However, when dealing with electrically conducting fluids, including plasmas undergoing the effects of electromagnetic fields, the hydrodynamic model should be replaced by appropriate physical-mathematical frameworks.
Of these, one of the simplest is MHD, which describes the behavior of plasmas under the influence of magnetic fields\cite{powell:1999}. For example, in \cite{grigoriadis:2010} the authors use the immersed boundary method to address the case of a MHD fluid interacting with a circular cilinder.

In Astrophysics one may cite as examples of MHD studies of interactions of fluids and bodies the paper \cite{johnstone:2015}, where the authors use the 3D code Nurgush to simulate the shocks between the winds from two low-mass stars forming a binary system, and \cite{vernisse:2013}, in which it is used the code A.I.K.E.F. (Adaptive Ion-Kinetic-Electron-Fluid) as a tool to treat lunar type plasma interactions.
Other pertinent examples include the study of exoplanets under the influence of the environment produced by their host stars: \cite{cohen:2015} addresses to Venus-like, non-magnetized exoplanets interacting with the wind from a M-dwarf star; \cite{nichols:2016} discusses exoplanets with magnetospheres undergoing Earth-like magnetospheric interaction with the solar wind; and in \cite{bourrier:2016} the authors analyze observations of the ``warm Neptune'' $\mbox{GJ}436\mbox{b}$ and the interaction between its exhosfere with the stellar wind. Further, it is worth mentioning \cite{spreiter:1970}, which focuses on the interaction of the solar wind with non-magnetized planets, while \cite{dryer:1973} studies the flow of the solar wind around Jupiter, Saturn, Uranus, Neptune and Pluto.

There are several computational schemes created to handle hydrodynamic and MHD problems. 
In this paper we use the FLASH code of the University of Chicago. 
However, it is important to point out that the tool which permits the placement of bodies in the simulations are, until this time, implemented and tested in such a code for pure hydrodynamic cases only.

Our aim here is to simulate the MHD interaction of the wind produced by a sun-like star with a non-magnetized planet, which has the approximate size of Earth and is placed at an orbital distance equal to the mean radius from the sun to Mercury. Furthermore, in our model the planet has no significant atmosphere. We achieve this by adapting the existing tools for simulating solid objects in pure hydrodynamic scenarios present in FLASH.

We investigate the influence of the viscosity on the regions around the planet, particularly its effects on the recirculation patterns and the behavior of the wake. Besides, in order to analyze the consistence of our scheme, we pay special attention to the magnetic field profiles and the mesh refinement in the MHD scenarios. For the sake of comparison, we perform a similar simulation in a pure hydrodynamic scenario.

The scheme presented here is interesting once it creates new perspectives for using the FLASH code, concerning the simulations of interactions of MHD fluids with rigid bodies. In addition, with the exponential growth of interest in research associated with exoplanets in the last two decades, both in observational and theoretical aspects, there is now strong interest in the studies of orbital evolution of planets due to their interaction with the protoplanetary disc, the central star and other planets (see, e.g., the recent work \cite{alvarado-gomez:2016}). Such studies are situated in a step that can be immediately extended from the work presented here.

This paper is organized as follows:
in Section~\ref{MHD} we show the basic formalism of MHD;
in Section~\ref{numerical} we discuss the numerical details of the simulations, concerning both the computational and the physical aspects;
in Section~\ref{results} the results and their respective discussions are presented, while the conclusions are given in Section \ref{conclusions}.

\section{Basic formalism of MHD}
\label{MHD}

Magnetohydrodynamics is one of the simplest frameworks for modelling the interaction between a conducting fluid and a magnetic field\cite{bateman:1978} and describes the macroscopic behavior of electrically conducting fluids, of which the most common is the plasma\cite{biskamp:2003}. Roughly speaking, MHD consists in the combination of the equations governing the fluid dynamics with Maxwell's equations of the electromagnetism.

Though the resulting system of equations can be presented in different ways, it is usually written in conservative form such that, in a fixed frame of reference (or Eulerian coordinate system), it assumes the form for the case where the viscosity is non-negligible:\cite{goedbloed:2004,lifschitz:1989}

\begin{align}
&\frac{\partial}{\partial t}(\rho\mathbf{v})= \nonumber \\
&\nabla\cdot\left[-\rho\mathbf{v}\mathbf{v}+\frac{1}{\mu}\mathbf{B}\mathbf{B}-\mathbb{I}\left(p+\frac{B^{2}}{2\mu}\right)\right] + \rho\nu\nabla^{2}\mathbf{v}, \label{mhd:1} \\[8pt]
&\frac{\partial\mathbf{B}}{\partial t}=\nabla\cdot(\mathbf{v}\mathbf{B}-\mathbf{B}\mathbf{v}), \label{mhd:2} \\[8pt]
&\frac{\partial\rho}{\partial t}=-\nabla\cdot(\rho\mathbf{v}), \label{mhd:3} \\[8pt]
&\frac{\partial\epsilon}{\partial t}=\nabla\cdot\left[-\left(\epsilon+p+\frac{\mathbf{B}^{2}}{2\mu}\right)\mathbf{v}+\frac{1}{\mu}(\mathbf{B}\cdot\mathbf{v})\mathbf{B}\right], \label{mhd:4} \\[8pt]
& \nabla\cdot\mathbf{B}=0, \label{mhd:5}
\end{align}
where: $\epsilon=\rho\mathbf{v}^{2}/2+p/(\gamma-1)+\mathbf{B}^{2}/2\mu$ is the total energy density of the fluid; $\mu$, $\mathbf{v}$, $\mathbf{B}$, $\rho$, $p$ are the magnetic permeability, the velocity, the magnetic field, the density and the pressure of the plasma; $\mathbb{I}$ is the $3\times 3$ identity matrix and $\nu$ is the kinematic viscosity\cite{goedbloed:2004}. Besides, it is considered a equation of state in the form $p=(\gamma-1)\epsilon$ where $\gamma$ is the adiabatic index.

From the form of such equations one may note that, with the exception of the term proportional to $\nu$ in Eq.~(\ref{mhd:1}), their right-hand sides are the divergent of the fluxes through the boundaries of the volume considered\cite{bateman:1978} and they represent, from top to bottom, the time evolution of the momentum, magnetic field, mass density and total energy, while Eq.~(\ref{mhd:5}) is the zero-divergence constraint on the magnetic field.

Computational codes use the conservative equations of MHD as shown here, once that particular form make them suitable for working out finite difference schemes. On the other hand, properties of the analytic equations can be used to validate the performance of numerical schemes\cite{bateman:1978}.

\section{Numerical Aspects}
\label{numerical}

The FLASH code has been originally developed for simulating astrophysical phenomena involving MHD and is distributed by the Center for Astrophysical Thermonuclear Flashes (FLASH Center) of the University of Chicago\footnote{\url{http://flash.uchicago.edu/site/flashcode}}. The default package used by this code for handling the adaptive-mesh refinement grid is PARAMESH\cite{macneice:2000}, which employs a refinement criteria adapted from L\"{o}hner's error estimator\cite{lohner:1987} with a threshold $10^{-2}$ in order to trigger the mesh refinement process. Besides, FLASH uses the Message-Passing Interface (MPI) library and HDF5 to allow portability on a variety of computers when dealing with parallel computation\cite{fryxell:2000}. Its modular architecture is such that it permits customization of the codes in order to simulate particular cases by means of changes in the algorithms and creation of new physics modules.

We consider five, six and seven levels of refinement in our simulations. However, we focus our analysis on the scenarios with five and seven levels; a result with six levels was generated just in order to investigate the convergence of the solutions and it is briefly mentioned in Subsection~\ref{BxDiffZero} (with a panel shown in Subsection~\ref{comparison}). Increasing the level of refinement by one duplicates the number of blocks in each coordinate and, as we start with a domain of $3\times 3\times 3$ blocks, we obtain $48\times 48\times 48$, $96\times 96\times 96$ and $192\times 192\times 192$ blocks with five, six and seven levels, respectively. Each block has $8\times 8\times 8$ cells.

In our MHD simulation we use the unsplit staggered mesh (USM) algorithm in order to solve Eqs.~(\ref{mhd:1})-(\ref{mhd:4}). It is based on the Godunov method, basically consisting in a conservative finite-volume scheme using spatial discretisation to solve the partial differential equations. For the pure hydrodynamic scenario it is used the unsplit hydro solver (UHS) which, in the present context, can be treated as a simplified version of USM where a fundamental difference is the presence of magnetic and electric fields in the latter. It is worth recalling that UHS uses the zone-edge data-extrapolated method as a specific predictor-corrector formulation.

The FLASH code employs as default the Roe approximate Riemann solver\cite{roe:1981}, which has been applied to a wide range of physical problems. However, despite the sucess of that solver, it can fail in regions of very low densities, producing unphysical states near strong rarefaction regions. Such a characteristic can represent a critical disadvantage in MHD scenarios, where in general the gas pressure is much less than the magnetic pressure\cite{li:2005}. Besides, due to the fact that the Roe solver demands eigen decomposition, it may become computationally costly in MHD problems. In order to overcome the mentioned limitations we use the HLL (Harten-Lax-van Leer-Contact) solver in both MHD and pure hydro scenarios, once this scheme satisfies the integral form of the conservation laws and it is computationally more robust\cite{einfeldt:1991}.

The time advancement of the equations in USM and UHS is based on a MUSCL-Hancock\cite{vanleer:1984} type algorithm and the code uses the constrained transport method to assure numerically the physical constraint given by Eq.~(\ref{mhd:5}) (see \cite{lee:2009}). On the other hand, all the simulations use a Courant Friedrichs Lewy (CFL) condition\cite{courant:1967} of $0.8$ and have an adiabatic index $\gamma=5/3$.

\subsection{Physical parameters of the problem}
\label{physicalparam}

We created a scenario representing a planet with the approximate size of Earth orbiting a sun-like star and placed at an orbital distance equal to the mean radius from the sun to Mercury, namely, $\unit[0.39]{AU}\approx \unit[6.0\times 10^{12}]{cm}$. Such a planet is inserted as a sphere of radius $\unit[6\times 10^{8}]{cm}$ and center at $\unit[(6.0\times 10^{9},0,0)]{cm}$ in a rectangular box whose dimensions are: $x\in [0.0,18.0] \unit[\times 10^{9}]{cm}$,  $y\in [-9.0,9.0] \unit[\times 10^{9}]{cm}$, $z\in [-9.0,9.0] \unit[\times 10^{9}]{cm}$ for the scenarios presented in Subsection \ref{PureHydro} and \ref{BxEqualZero}; and $x\in [0.0,18.0] \unit[\times 10^{9}]{cm}$,  $y\in [-12.0,6.0] \unit[\times 10^{9}]{cm}$, $z\in [-12.0,6.0] \unit[\times 10^{9}]{cm}$ for the case shown in Subsection \ref{BxDiffZero}.

The magnetic field $\mathbf{B}$ to be used as an of the initial parameters in our simulations is determinated by means of Parker's model for the solar wind, such that its components are written in spherical coordinates as\cite{parker:1958}

\begin{align}
    &B_r = B_0\left(\frac{b}{r}\right)^{2}, \label{parkerB:1} \\
    &B_\theta = 0, \label{parkerB:2} \\
    &B_\phi = B_0\left(\frac{\Omega}{v_{\mbox{\scriptsize{sw}}}}\right)(r-b)\left(\frac{b}{r}\right)^{2}\sin\theta, \label{parkerB:3}
\end{align}
where $B_0$ and $b$ are constants, $\Omega$ is the angular velocity of the sun, $v_{\mbox{\scriptsize{sw}}}$ is the radial velocity of the solar wind and $r$ is the heliocentric distance. As we are assuming that the orbit of our hypothetical planet is in the ecliptic plane ($\theta=\pi/2$) and given the fact that Eqs.~(\ref{parkerB:1})-(\ref{parkerB:3}) do not depend on $\phi$, we can, for the sake of convenience, consider that $B_r$ and $B_\phi$ have the directions of the x-axis and y-axis in our domain, respectively.

Since, according to \cite{kivelson:1995}, the components of the magnetic field at \unit[1]{AU} (written as $B^e_r$ and $B^e_\phi$) have values such that $\sqrt{(B^e_r)^2+(B^e_\phi)^2}=\unit[7]{nT}$ and $B^e_\phi/B^e_r \approx 1$, we can use Eqs.~(\ref{parkerB:1})-(\ref{parkerB:3}) to evaluate such components at \unit[0.39]{AU} (writting them as $B^m_r$ and $B^m_\phi$ in this case). With effect, from Eq.~(\ref{parkerB:1}) we deduce that $B^m_r(0.39)^2=B^e_r(1)^2$, giving $B^m_r=\unit[32.5]{nT}$. Now, dividing Eq.~(\ref{parkerB:3}) by Eq.~(\ref{parkerB:1}) and using $v_{\mbox{\scriptsize{sw}}}=\unit[400]{km~s^{-1}}$, $\Omega=\unit[2.7\times 10^{-6}]{\mbox{rad}~s^{-1}}$ and $b=\unit[4.6\times 10^{-2}]{AU}$ \cite{tautz:2011}, we have $B_\phi/B_r\approx r$; for $r=\unit[0.39]{AU}$ we obtain finally $B^m_\phi=\unit[12.7]{nT}$.

From Eqs.~(\ref{parkerB:1})-(\ref{parkerB:3}) we note that Parker's model does not define a component perpendicular to $B_r$ and $B_\phi$. On the other hand, the presence of a $B_\theta$ different from zero is justified, for example, by the transport of magnetic fields on the solar surface and turbulence\cite{korth:2011}, making interesting the inclusion of such a component in our scenarios. According to \cite{korth:2011}, measurements of $B_\theta$ taken between $\unit[0.31]{AU}$ and $\unit[0.47]{AU}$ by spacecrafts such as MESSENGER and Helios present large flutuations around zero, making difficult in principle to choose a ``typical'' value to be used here. However, as we can deduce from the histograms shown in \cite{korth:2011}, more than $\approx 90 \%$ of the pertinent observational data lie in the interval $\approx [-15,15]\unit{nT}$, suggesting us that it would be reasonable to consider an initial $B_\theta$ (written as $B_z$ hereafter) of $\sim\unit[10]{nT}$ in our simulations.

The remaining initial parameters, namely, $\rho$ (obtained from the proton density $n_p$ and the electron density $n_e$) and $p$ at $r=\unit[0.39]{AU}$ can be obtained by a similar procedure to the one used in the evaluation of $B^m_r$ and $B^m_\phi$. With effect, from Parker's model, we may consider that $n_e$ and $n_p$ has a dependence on $r$ in the form $n_{e,p} \propto r^{-2}$\cite{parker:1958}. Besides, let us assume that the proton temperature $T_p$ and the electron temperature $T_e$ vary with $r$ as $T_p \propto r^{-1}$ and $T_e \propto r^{-1/2}$\cite{kivelson:1995}. Now, from the fact that at $r=\unit[1]{AU}$ we have $n_p=n_e=\unit[7]{cm^{-3}}$, $T_p=\unit[1.2\times 10^{5}]{K}$ and $T_e=\unit[1.4\times 10^{5}]{K}$\cite{kivelson:1995}, we are able to deduce that such variables have the values $n=\unit[46]{cm^{-3}}$ (dropping the subscripts), $T_p=\unit[3.08\times 10^{5}]{K}$ and $T_e=\unit[2.24\times 10^{5}]{K}$ at $r=\unit[0.39]{AU}$.

The pressure is calculated by $p=nk_B(T_p+T_e)$ where $k_B$ is the Boltzmann constant, giving $p=\unit[3.38\times 10^{-9}]{dyn~cm^{-2}}$; besides, $\rho=n(m_p+m_e)$ with $m_p$ and $m_e$ representing the proton and electron masses, yielding $\rho=\unit[1.17\times 10^{-23}]{g~cm^{-3}}$. 

The values of $\rho$, $p$ and $\mathbf{B}$ calculated above are used as initial conditions of the domain (inside the planet we use different conditions, as explained subsequently). On the other hand, the initial $\mathbf{v}$ of the domain is given by $(v_{\mbox{\scriptsize{sw}}},v_{\phi},0)$. From Parker's model we have $v_{\phi}=\Omega(r-b)\sin\theta$ which, in our case, yields the value $v_{\phi}=\unit[140]{km~s^{-1}}$. Table~\ref{tab:1} summarizes the initial $\rho$, $p$, $\mathbf{v}$ and $\mathbf{B}$ to be used in the simulations.

\begin{table}
\centering
\caption{Numerical values of the initial parameters $\rho$, $p$, $\mathbf{v}$ and $\mathbf{B}$ used in the domain and in the defined boundary condition (which represents the stellar wind flowing from $x=0$).}
\label{tab:1}
\begin{tabular}{cccc}
\hline\noalign{\smallskip}
$\rho$ & $p$ & $\mathbf{v}$ & $\mathbf{B}$ \\
($\times 10^{-23}$) & ($\times 10^{-9}$) & ($\times 10^{7}$) & \\
$\unit{g~cm^{-3}}$ & $\unit{dyn~cm^{-2}}$ & $\unit{cm~s^{-1}}$ & $\unit{nT}$ \\
\noalign{\smallskip}\hline\noalign{\smallskip}
1.17 & 3.38 & (4.0,1.4,0) & (32.5,12.7,10.0) \\
\noalign{\smallskip}\hline
\end{tabular}
\end{table}

The stellar wind is represented as flowing from the border at $x=0$ of the domain with the velocity given by Table~\ref{tab:1}. In order to do so we employ the user defined boundary condition, defining at such a border the values for $\rho$, $p$, $\mathbf{v}$ and $\mathbf{B}$ given in Table~\ref{tab:1}. The outflow boundary condition, which stands for a zero normal gradient at the region being considered, is applied to the remaining edges. As a particular case shown in Appendix~1, we performed a simulation where we consider the user defined condition at the left ($x=0$), top and bottom boundaries, whereas at the right one we maintain the outflow condition.

The physical initial conditions inside the solid body are defined in the following way: $\mathbf{v}_{\mbox{\scriptsize{body}}}=0$, $\mathbf{B}_{\mbox{\scriptsize{body}}}=0$, $\rho_{\mbox{\scriptsize{body}}}=\unit[1.17\times 10^{-22}]{g~cm^{-3}}$ and $p_{\mbox{\scriptsize{body}}}=\unit[3.38\times 10^{-9}]{dyn~cm^{-2}}$. Actually, in preliminar simulations we tested different values for $\rho_{\mbox{\scriptsize{body}}}$ and $p_{\mbox{\scriptsize{body}}}$ and we verified that the results are not noticeably affected by the exact numerical choice of such parameters in the cases where they are greater than or equal to, respectively, $\rho$ and $p$ in Table~\ref{tab:1}. Despite the fact that in a typical planet $\rho\sim\unit[1]{g~cm^{-3}}$, we consider the mentioned value of $\rho_{\mbox{\scriptsize{body}}}$ for the sake of convenience in the treatment and visualization of the results. It is worth noting that inside rigid bodies the MHD equations do not evolve; further, the code applies the reflecting boundary condition at the surface of such objects.

\subsection{Values of the viscosity to be used in the simulations}
\label{viscosity}

According to the model for the kinematic viscosity $\nu$ of the solar wind discussed in \cite{subramanian:2012}, we have $\nu=\unit[300]{km^2~s^{-1}}$ at $r=\unit[0.39]{AU}$, giving us one of the values of $\nu$ to be employed in our scenarios. On the other hand, \cite{tejada:2005} presents a estimate of $\nu\sim\unit[1000]{km^2~s^{-1}}$ at $r=\unit[0.72]{AU}$ (the heliocentric distance of Venus), while the model by \cite{subramanian:2012} yields $\nu\approx\unit[600]{km^2~s^{-1}}$ at the same $r$. Such a fact suggests us that, according to the literature, there may be discordance about the evaluations of $\nu$ corresponding to each $r$. Therefore, besides considering $\nu=\unit[300]{km^2~s^{-1}}$, it would be interesting to simulate additional cases with different values of $\nu$. For this purpose we use $\nu=\unit[1000]{km^2~s^{-1}}$ and $\nu=\unit[5000]{km^2~s^{-1}}$. The latter is artificially high and was included in order to analyze the effects of the viscosity on the processes being simulated.

It is useful to define the Reynolds number $Re$, which may be written in function of $\nu$ as\cite{grigoriadis:2010}

\begin{equation}
\label{reynolds}
Re=\frac{uD}{\nu},
\end{equation}
where $u$ is the velocity of the fluid ($\sqrt{v^{2}_{\mbox{\scriptsize{sw}}}+v^{2}_{\phi}}$ in our case) and $D$ represents a characteristic linear dimension of the body. Here, $D$ is considered as the diameter of the planet.


The model used in our simulations is essentially collisionless, once in such a formalism the viscosity is considered as totally caused by protons being scattered by ``kinks'' in the magnetic fields, while the proton-proton collisions are neglected in the deductions. Though such a model is suitable for our purposes, the solar wind may in fact be weakly collisional for the scales used here; with effect, strictly speaking, the wind is considered collisionless up to $\sim\unit[10]{R_{\odot}}$. See \cite{marsch:2006} for a detailed discussion.

It is worth mentioning that in the regions where the solar wind is collisional, we have the predominance of Coulomb collisions. Such processes have influence on the physical characteristics of the plasma, such as affecting the ion velocity distributions. See \cite{livi:1986} for details.

\section{Results}
\label{results}

In this section we present the simulations for three cases: purely hydrodynamic, MHD with the initial $\mathbf{B}$ given by Table~\ref{tab:1} and MHD considering an initial $\mathbf{B}$ in the form $(0,12.7,10.0)~\unit{nT}$. 

\subsection{Purely hydrodynamic case}
\label{PureHydro}

\begin{figure*}
\centering
\begin{tabular}{cc}
\includegraphics[width=0.47\linewidth]{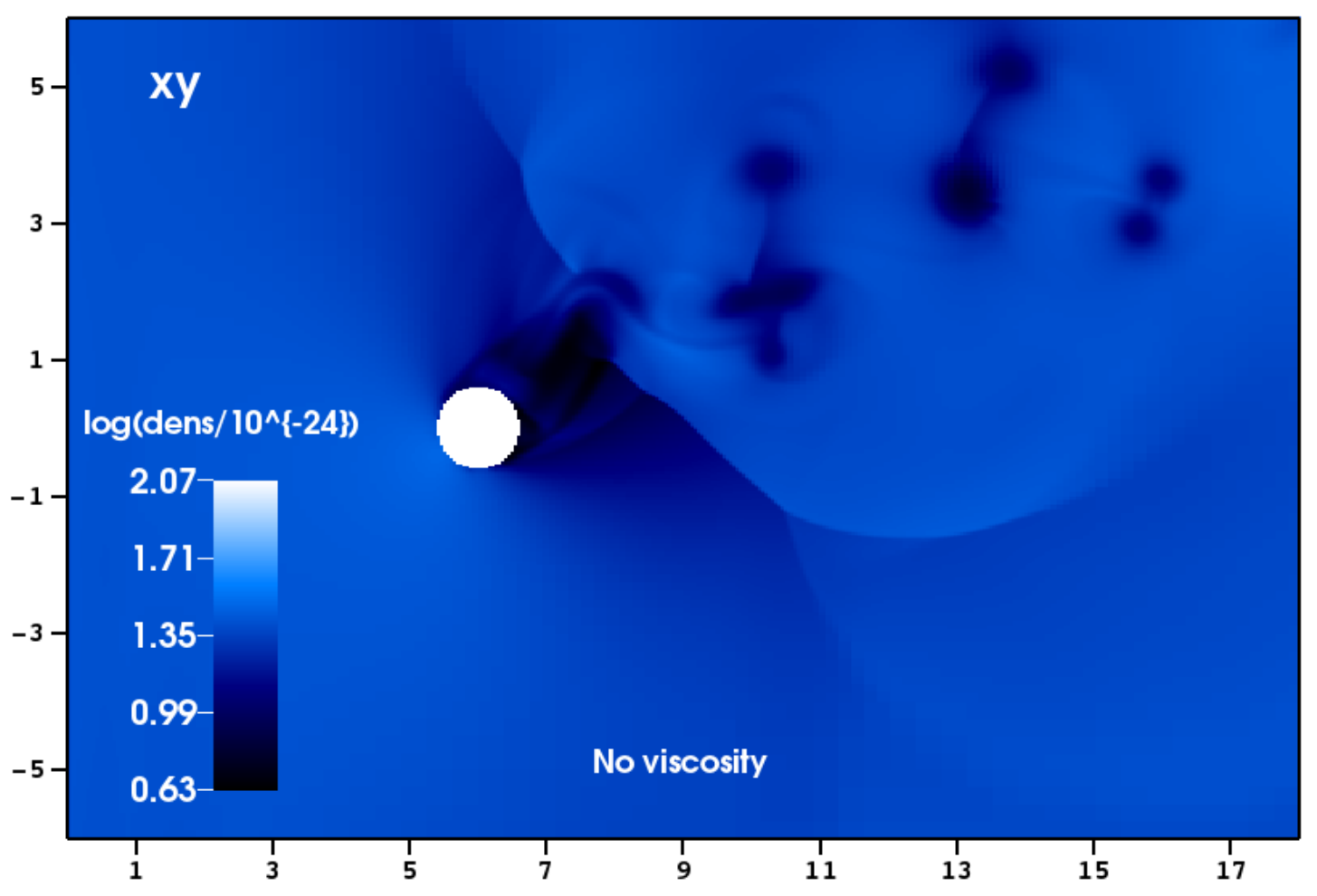}
\includegraphics[width=0.47\linewidth]{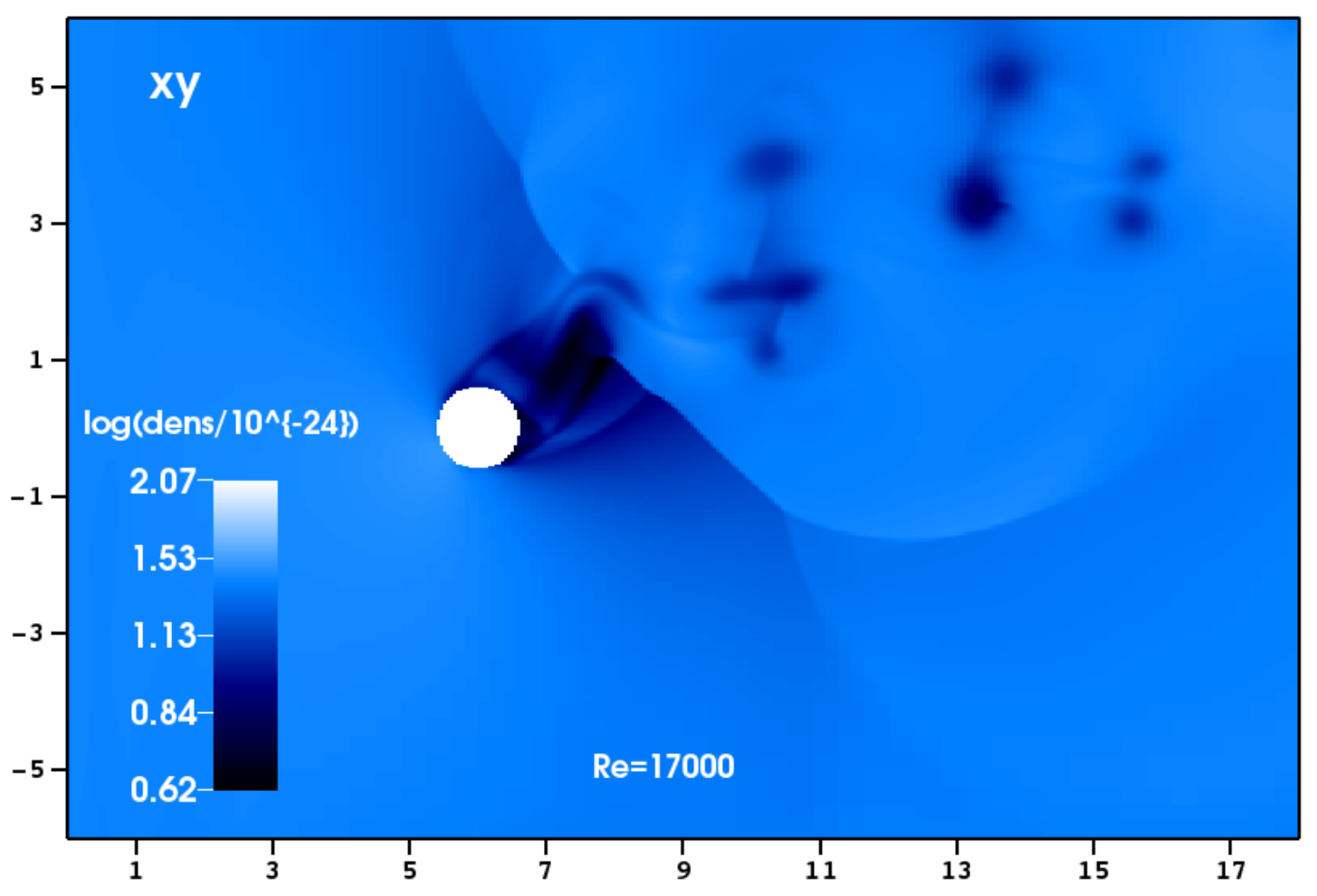}\\
\end{tabular}
\caption{Density profiles for the purely hydrodynamic simulations at $t=\unit[1200]{s}$ in the xy-plane. The density is given in $\log(\rho/\unit[10^{-24}]{g~cm^{-3}})$ and the dimensions of the box are in $\unit[10^{9}]{cm}$. Right: scenario with no viscosity; left: scenario with $\nu=\unit[300]{km^2~s^{-1}}$ ($Re=17000$).}
\label{figs:1}
\end{figure*}

Figure~\ref{figs:1} shows the density profiles in the xy-plane and at the instant $t=\unit[1200]{s}$ (after the vanishing of the transients present at the initial instants of the simulation) for the purely hydrodynamic case. We considered the values of $\rho$, $p$ and $\mathbf{v}$ in Table~\ref{tab:1} as initial parameters of the domain; besides, we used five levels of refinement. The dimensions of the box are in $\unit[10^{9}]{cm}$ and $\rho$ is in units of $\log(\rho/\unit[10^{-24}]{g~cm^{-3}})$.

The left profile in Fig.~\ref{figs:1} represents the case where we neglect the viscosity; the right one corresponds to $\nu=\unit[300]{km^2~s^{-1}}$ ($Re=17000$). We may note the formation of vortices past the planet, which tend to the right top region of the domain due to the presence of $v_{\phi}$. Comparing both panels we note that the differences between their correspondent patterns caused by the viscosity are very small for that particular value of $\nu$. Also, both scenarios are characterized by $\rho\approx\unit[3.0\times10^{-23}]{g~cm^{-3}}$ and $p\approx\unit[2.0\times 10^{-8}]{dyn~cm^{-2}}$ at the left side of the body and $\rho\approx\unit[1.0\times10^{-23}]{g~cm^{-3}}$ and $p\approx\unit[3.0\times 10^{-9}]{dyn~cm^{-2}}$ at right (in the wake between $x=\unit[7\times 10^{9}]{cm}$ and $x=\unit[9\times 10^{9}]{cm}$). 

It is worth noting the shock seen between the mentioned wake and the vortices, where its left side is characterized by $\rho=\unit[1.3\times 10^{-23}]{g~cm^{-3}}$, $p=\unit[4\times 10^{-9}]{dyn~cm^{-2}}$ and $|\mathbf{v}|=\unit[3.5\times 10^{7}]{cm~s^{-1}}$, while the right one has $\rho=\unit[2.0\times 10^{-23}]{g~cm^{-3}}$, $p=\unit[1.5\times 10^{-8}]{dyn~cm^{-2}}$ and $|\mathbf{v}|=\unit[2.0\times 10^{7}]{cm~s^{-1}}$.


\begin{figure*}
\centering
\begin{tabular}{cc}
\includegraphics[width=0.35\linewidth]{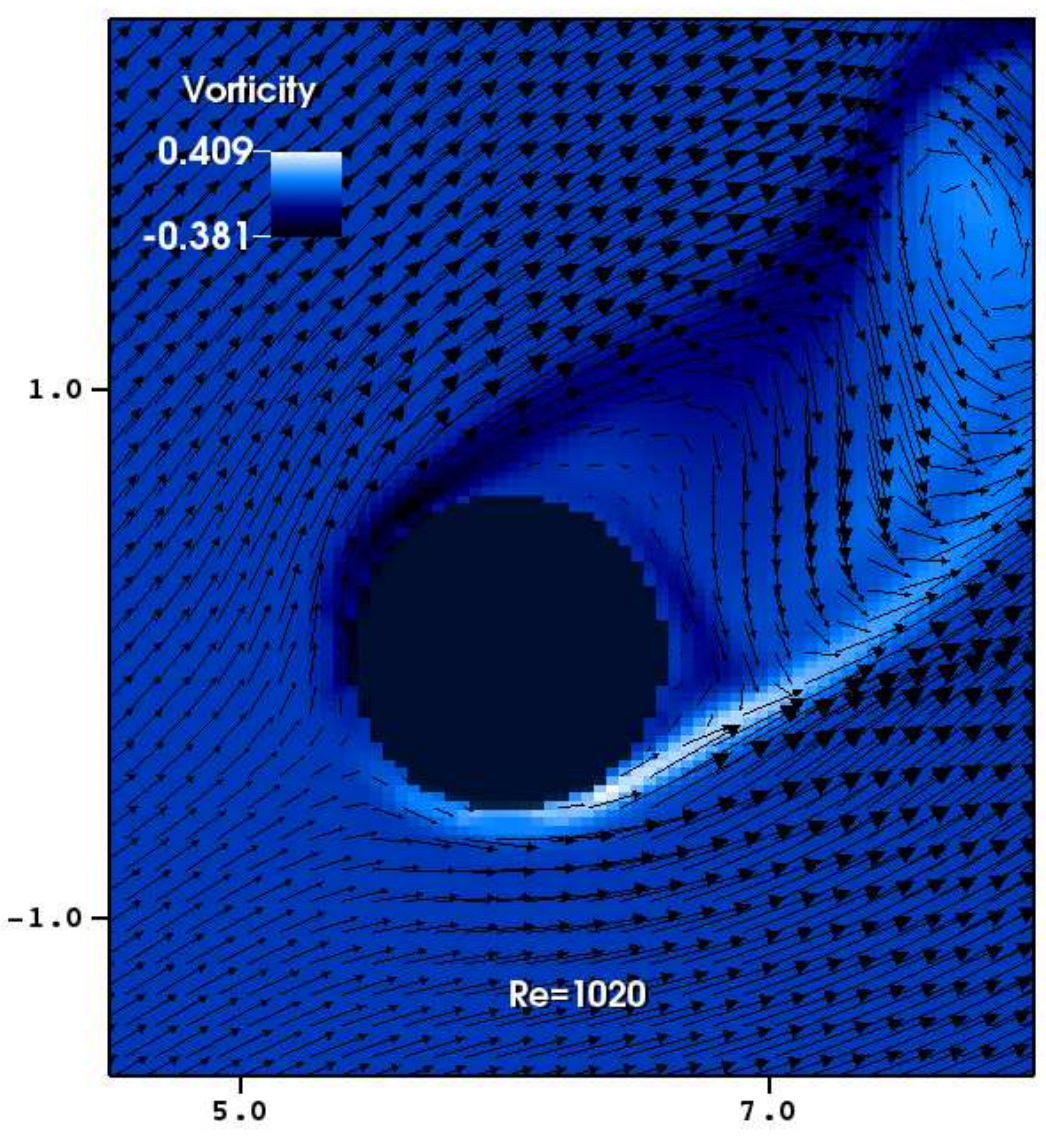}
\includegraphics[width=0.35\linewidth]{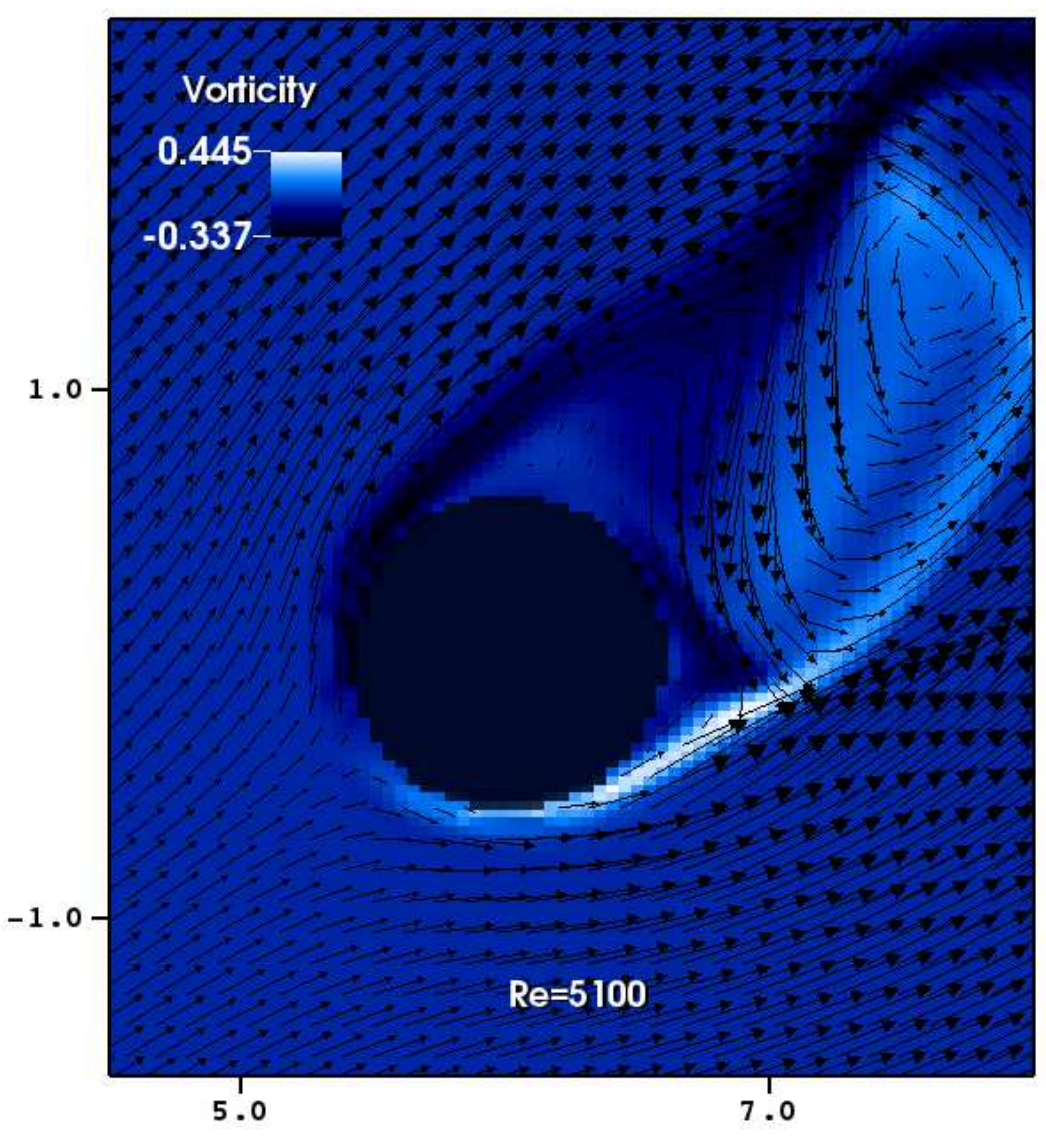}\\
\includegraphics[width=0.35\linewidth]{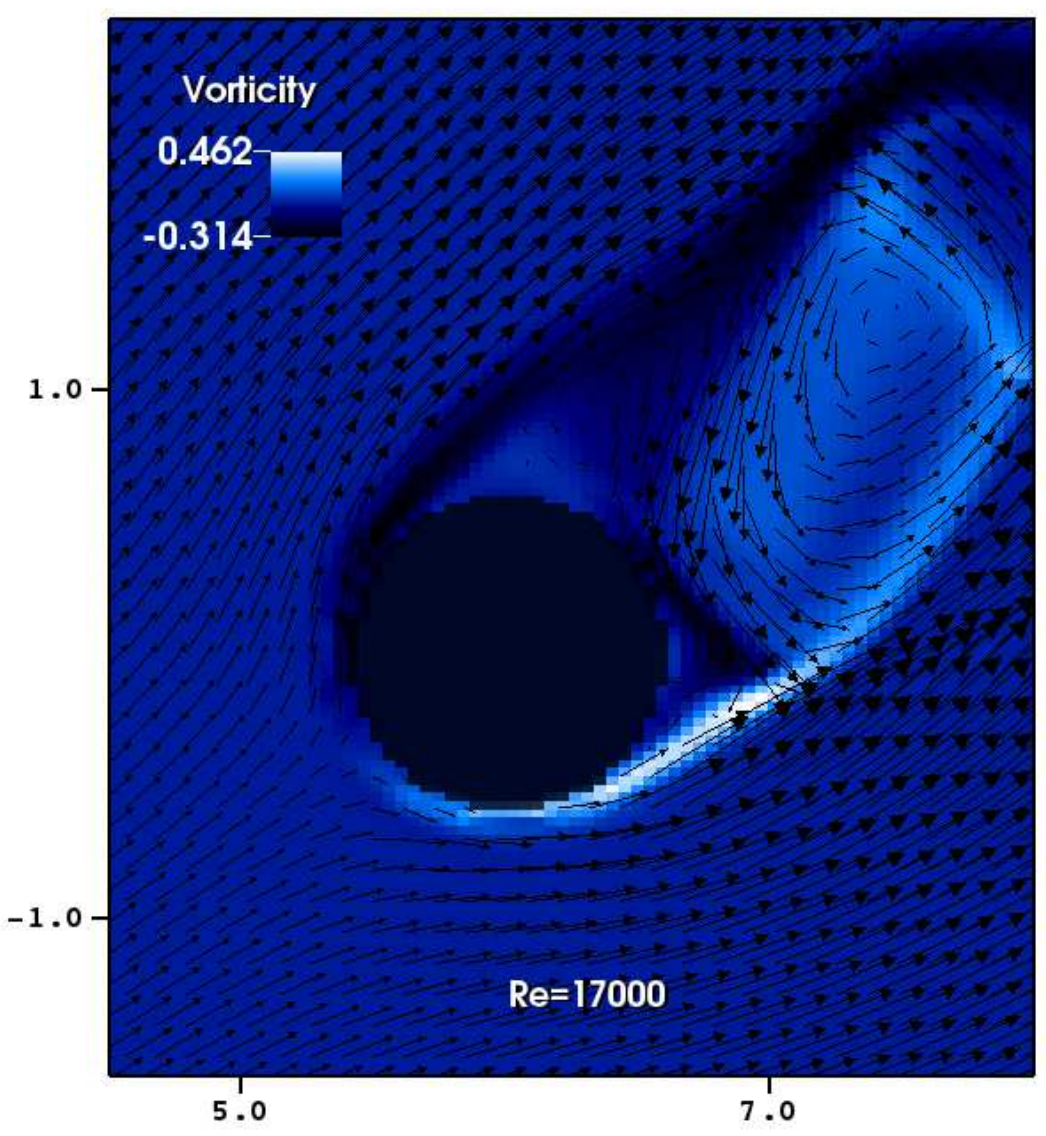}
\includegraphics[width=0.35\linewidth]{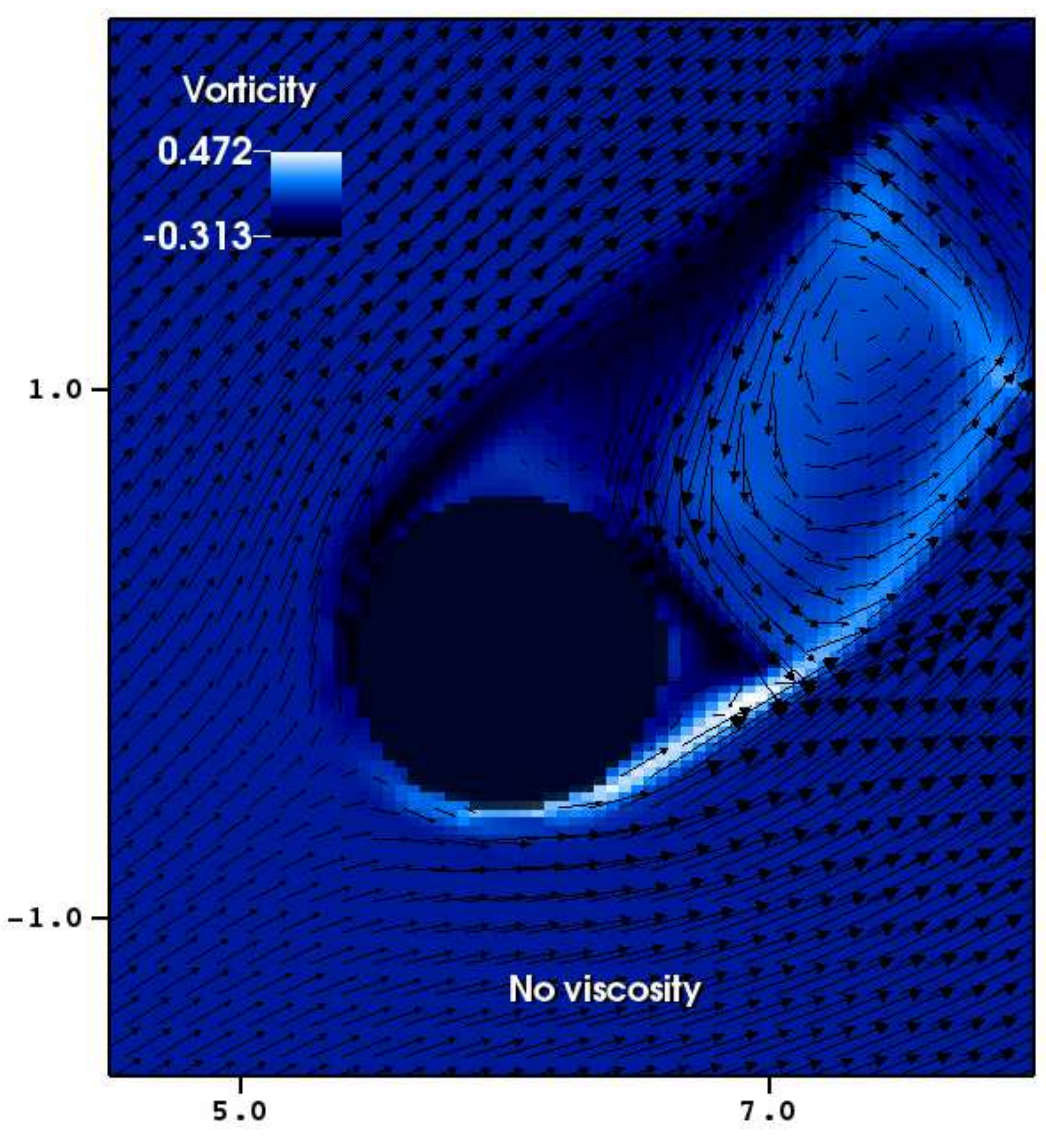}\\
\end{tabular}
\caption{Velocity vector field and vorticity profiles for the hydrodynamic simulations in the xy-plane and at $t=\unit[1200]{s}$. From top to bottom and from left to right: scenarios with $\nu=\unit[5000]{km^2~s^{-1}}$ ($Re=1020$), $\nu=\unit[1000]{km^2~s^{-1}}$ ($Re=5100$), $\nu=\unit[300]{km^2~s^{-1}}$ ($Re=17000$) and with no viscosity. The maximum $|\mathbf{v}|$ are of $\approx\unit[5.0\times 10^7]{cm~s^{-1}}$ and the dimensions of the box are in $\unit[10^{9}]{cm}$.}
\label{figs:2}
\end{figure*}

Figure~\ref{figs:2} presents the velocity vector field and the vorticity profiles for the purely hydrodynamic simulations at $t=\unit[1200]{s}$ in the xy-plane. Note that we focus on the regions around the planet. The vorticity $\omega_z$ is calculated from $\mathbf{v}$ by means of

\begin{equation}
\label{vort}
\omega_z=\frac{\partial v_y}{\partial x}-\frac{\partial v_x}{\partial y}.
\end{equation}
Four scenarios are considered: using $\nu=\unit[5000]{km^2~s^{-1}}$ ($Re=1020$), $\nu=\unit[1000]{km^2~s^{-1}}$ ($Re=5100$), $\nu=\unit[300]{km^2~s^{-1}}$ ($Re=17000$) and with no viscosity. The maximum value of $|\mathbf{v}|$ in the four profiles of Fig.~\ref{figs:2} are of $\approx\unit[5.0\times 10^7]{cm~s^{-1}}$. The length $L$ of the reciculation region is calculated from the surface of the planet (point $x=\unit[6.6\times 10^{9}]{cm},y=0$) until the edge of the circulation pattern seen at the right top of panels of Fig.~\ref{figs:2}. For the range of values considered here, $L$ slightly decreases as $Re$ increases: for $Re=1020$, $Re=5100$ and $Re=17000$ we have $L=\unit[2.3\times 10^{9}]{cm}$, $L=\unit[2.1\times 10^{9}]{cm}$ and $L=\unit[2.0\times 10^{9}]{cm}$, respectively (see Fig.~\ref{figs:9}); besides, $L=\unit[1.8\times 10^{9}]{cm}$ with no viscosity. On the other hand, the maximum value of $|\omega_z|$ increases as $Re$ increases (see the values of $|\omega_z|$ for each case in Fig.~\ref{figs:2} and the diagrams in Fig.~\ref{figs:9}.) Note that, for convenience, $|\omega_z|$ for this case is shown multiplied by four in Fig.~\ref{figs:9}.

Our simulations might be compared to other results found in the literature. In fact, the particular case of the solar wind interacting with the Moon presented in \cite{spreiter:1970} has a characteristic in common with our hydrodynamic simulations: in both cases there is no formation of bow shock, once the authors of such a paper considered that the Moon has no magnetic field and ionosphere to deflect the solar wind. 

Observing Fig.~\ref{figs:2} we note that the fluid is deflected as it pass around the planet once that in FLASH the surface of rigid bodies is treated as a reflecting boundary. However, the velocities of the fluid drop nearly to zero at the region where it reaches radially the surface of the planet. On the other hand, we should bear in mind that in \cite{spreiter:1970} the particles of the solar plasma which hit the lunar surface are stopped and removed from the flow. Both scenarios are intrinsically different once in \cite{spreiter:1970} the fluid is absorbed by the surface of the body.

In \cite{grigoriadis:2010} it is shown, among other results, the influence of the viscosity on the size of the recirculation regions for a hydrodynamic fluid interacting with a cylinder. The authors found that, for $Re\gtrapprox50$, higher values of $Re$ are related to smaller $L$. We observed a similar behavior in our hydrodynamic simulations, though the geometry of the body in \cite{grigoriadis:2010} is not the same as the one used here (see Fig.~\ref{figs:9}).

\subsection{MHD scenario with initial $B_x=\unit[32.5]{nT}$}
\label{BxDiffZero}

\begin{figure*}
\centering
\begin{tabular}{cc}
\includegraphics[width=0.47\linewidth]{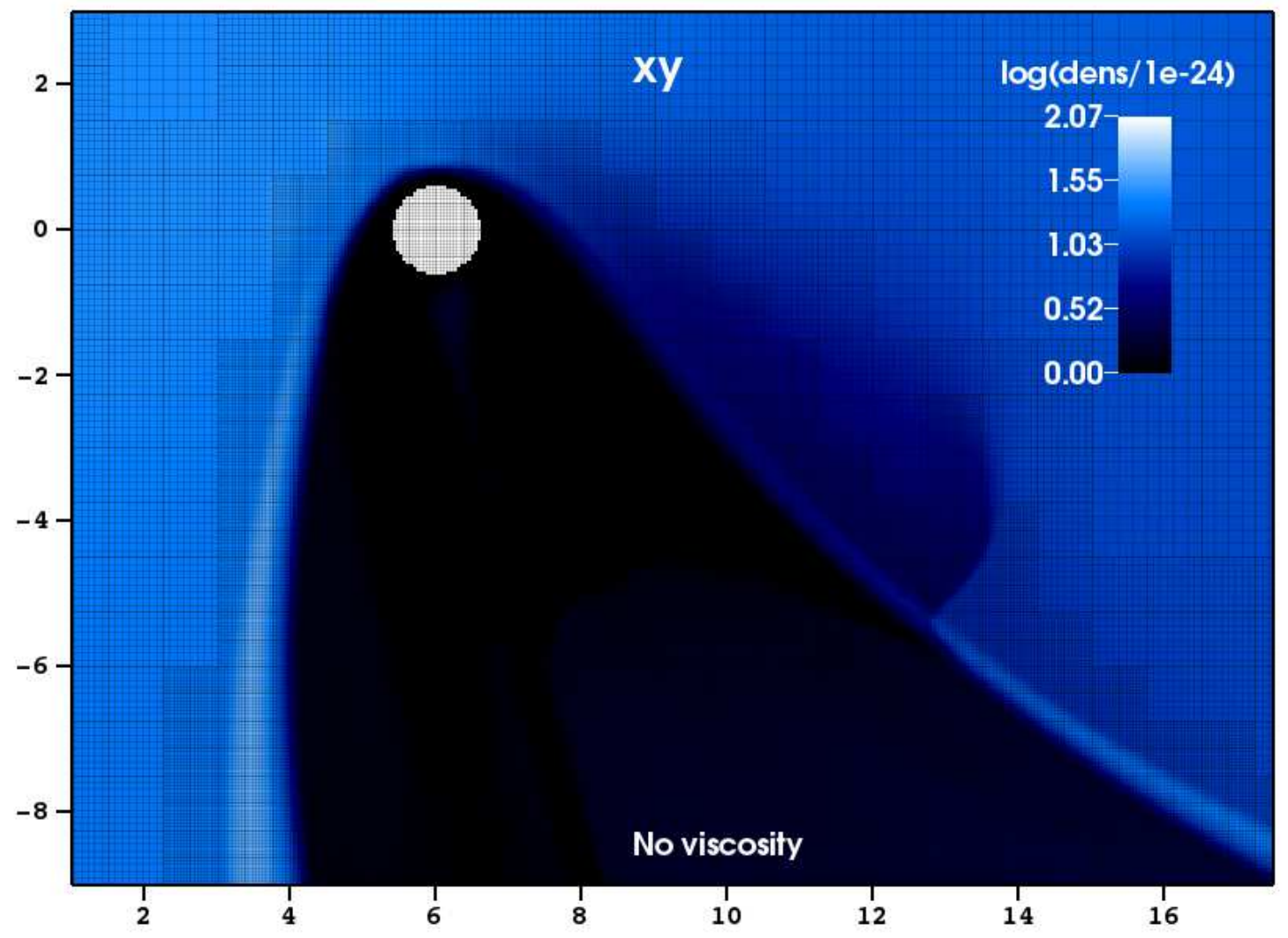}
\includegraphics[width=0.47\linewidth]{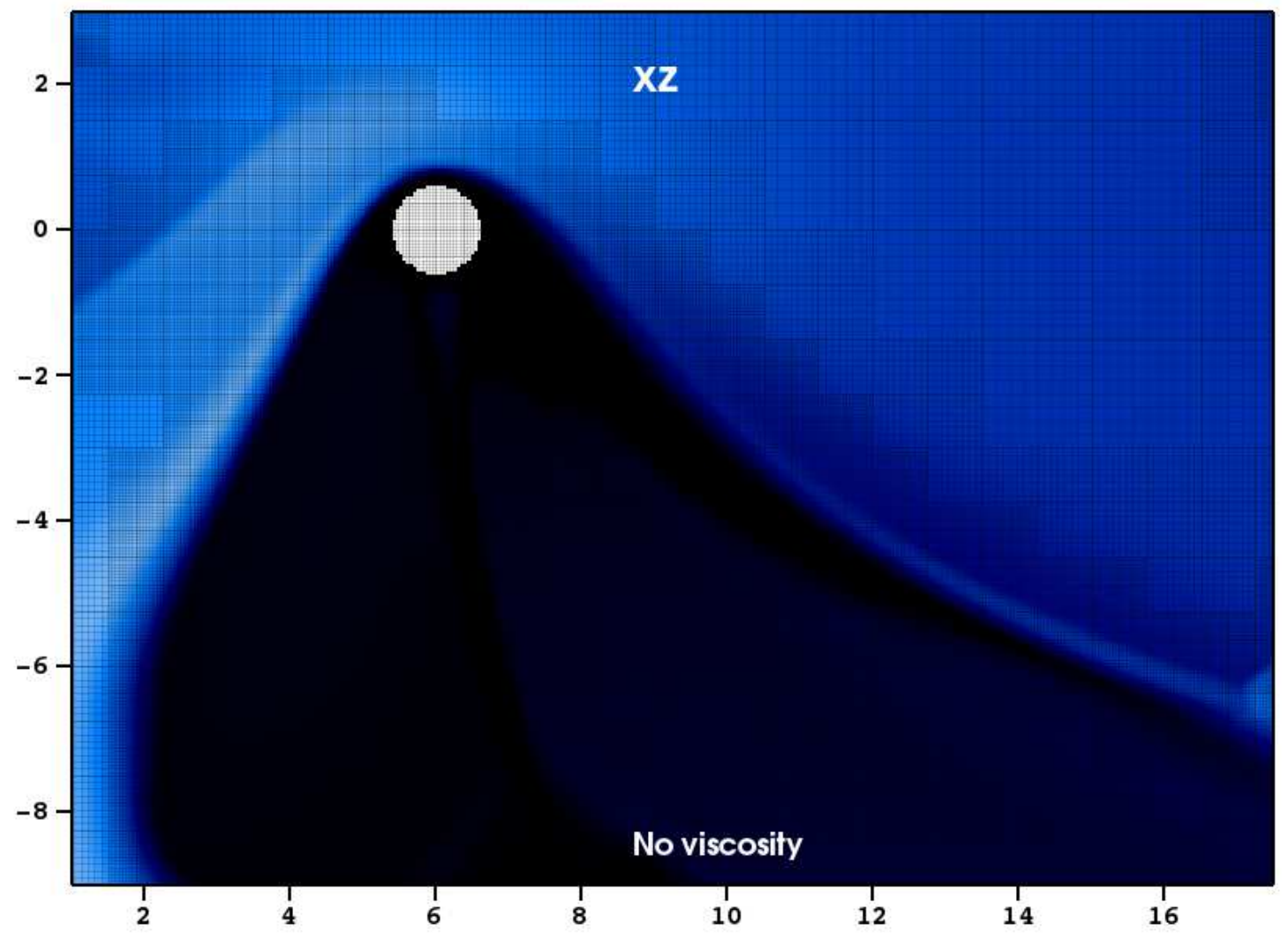}\\
\includegraphics[width=0.47\linewidth]{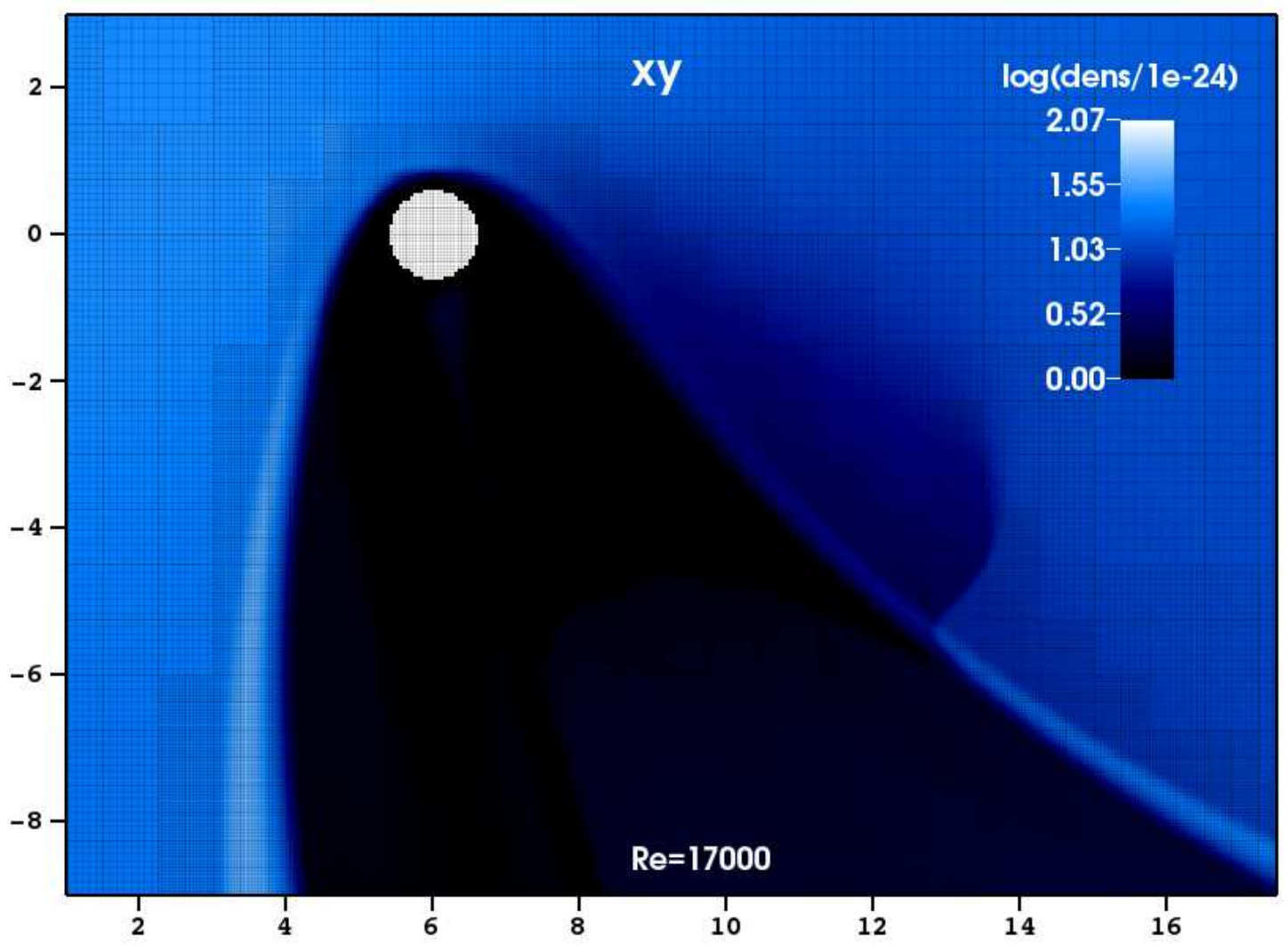}
\includegraphics[width=0.47\linewidth]{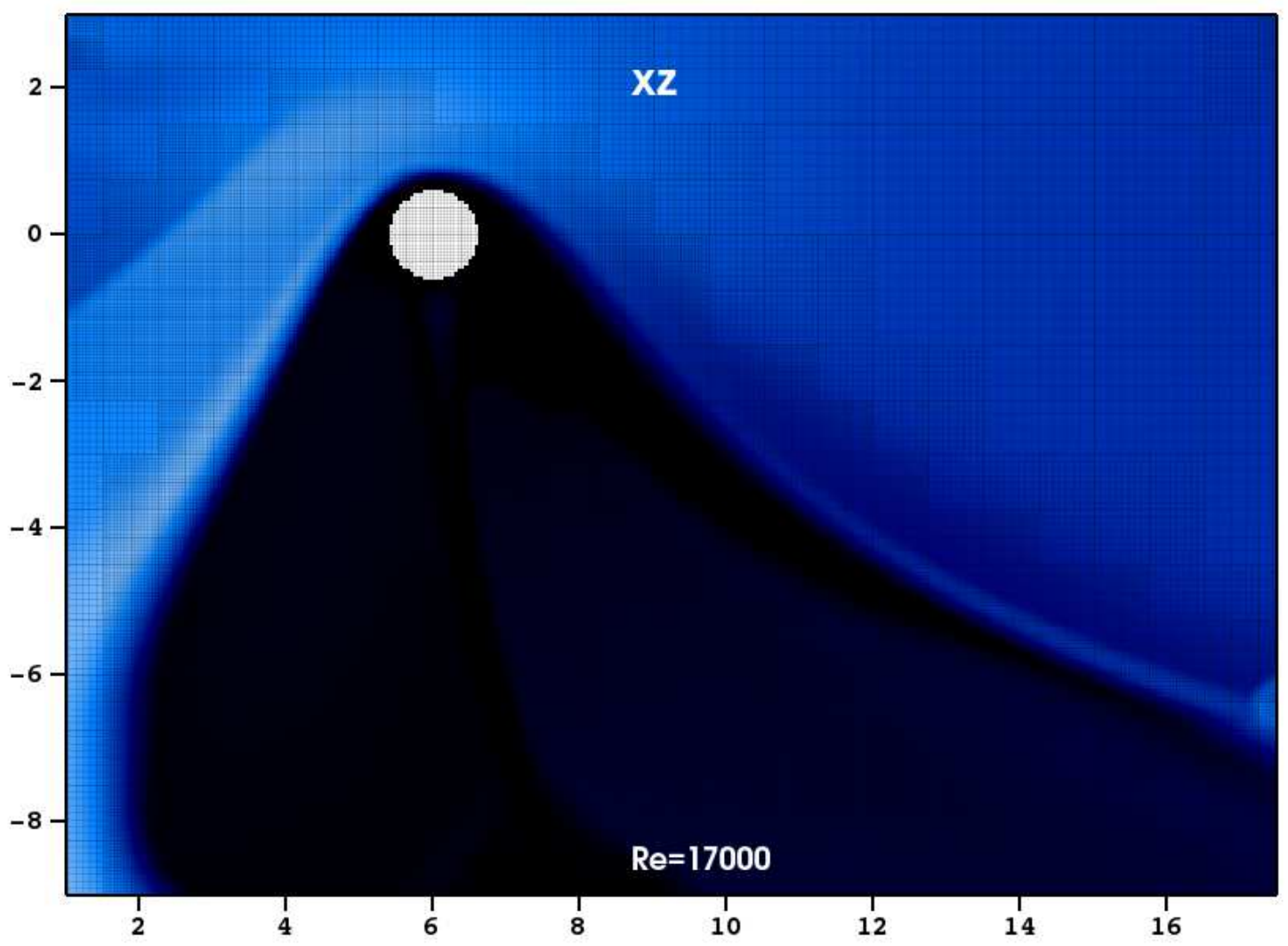}\\
\end{tabular}
\caption{Density profiles for the MHD simulations at $t=\unit[1400]{s}$ and using five levels of refinement. The density is given in $\log(\rho/\unit[10^{-24}]{g~cm^{-3}})$ and the dimensions of the box are in $\unit[10^{9}]{cm}$. Upper panels: xy and xz-planes for the scenario with no viscosity; lower panels: same as the upper ones but considering $\nu=\unit[300]{km^2~s^{-1}}$ ($Re=17000$).}
\label{figs:3}
\end{figure*}

Figure~\ref{figs:3} presents the density profiles of the MHD simulations at $t=\unit[1400]{s}$ in the xy-plane (left panels) and xz-plane (righ panels) with the initial conditions of the domain given in Table~\ref{tab:1} and using five levels of refinement; the upper and lower panels correspond respectively to the scenarios with no viscosity and considering $\nu=\unit[300]{km^2~s^{-1}}$ ($Re=17000$). Generally speaking, there is the formation of a thick, distinct bow shock with $\rho\approx\unit[3\times 10^{-23}]{g~cm^{-3}}$ and $p\sim\unit[1\times 10^{-8}]{dyn~cm^{-2}}$. The low-density tails are characterized by $\rho\sim\unit[10^{-24}]{g~cm^{-3}}$ and $p\sim\unit[1\times 10^{-7}]{dyn~cm^{-2}}$ at their central regions in both scenarios. Further, note that, for five levels of refinement, the viscosity has no noticeable effects on the density profiles.

\begin{figure*}
\centering
\begin{tabular}{cc}
\includegraphics[width=0.47\linewidth]{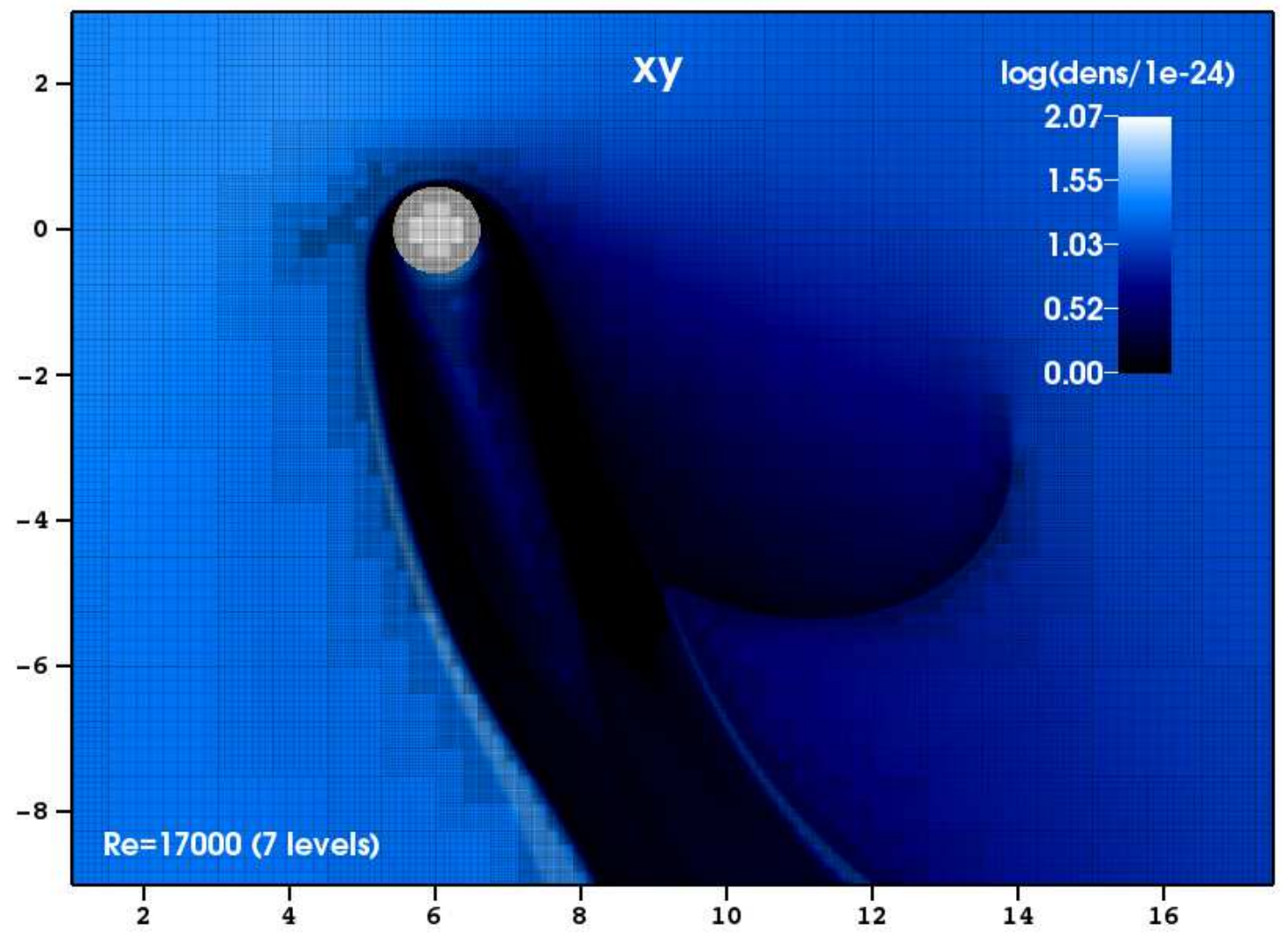}
\includegraphics[width=0.47\linewidth]{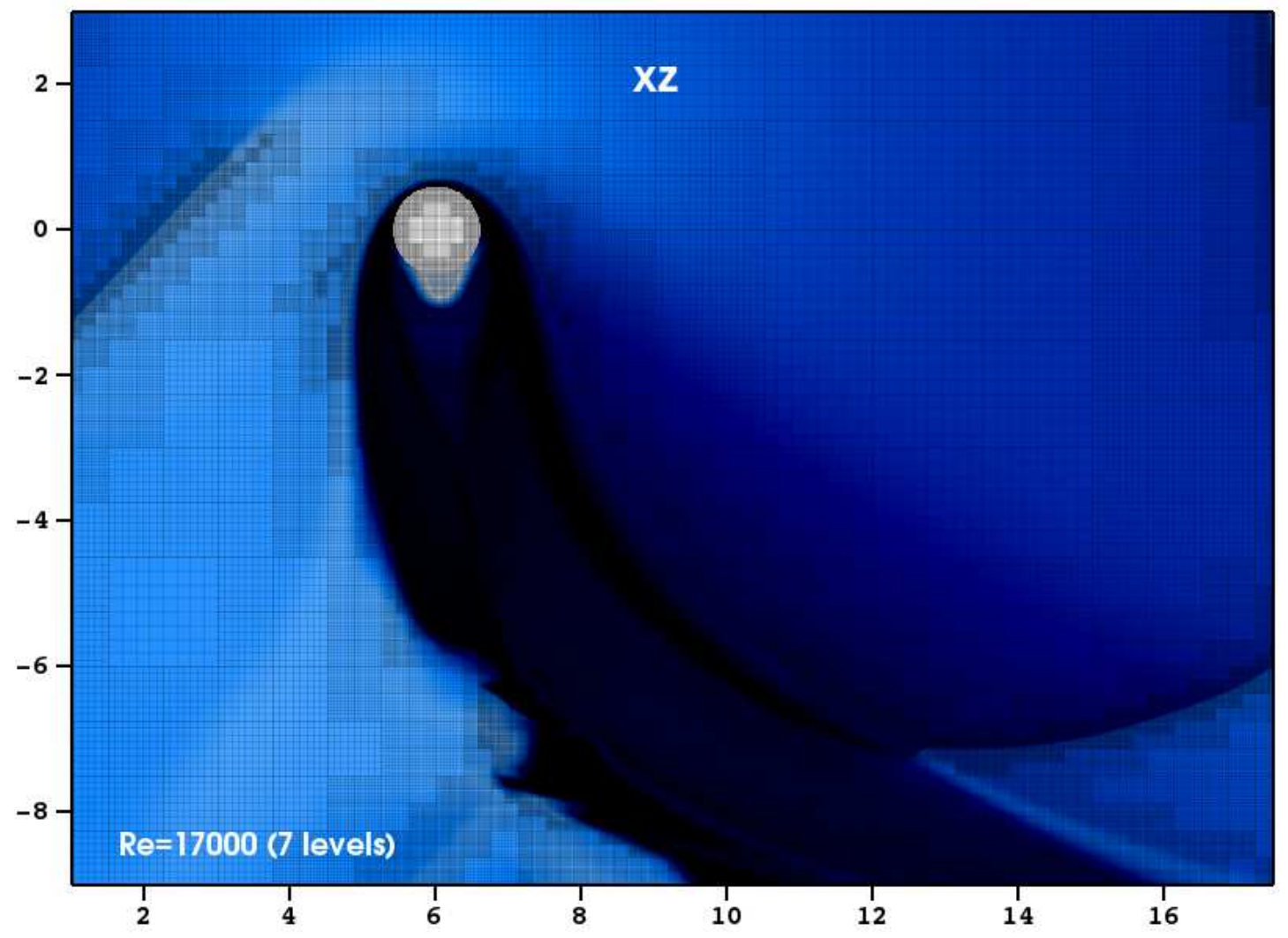}\\
\end{tabular}
\caption{Density profiles for the MHD simulations at $t=\unit[1400]{s}$. The density is given in $\log(\rho/\unit[10^{-24}]{g~cm^{-3}})$ and the dimensions of the box are in $\unit[10^{9}]{cm}$. The panels show the xy and xz-planes for the scenario with $\nu=\unit[300]{km^2~s^{-1}}$ ($Re=17000$) and considering seven levels of refinement.}
\label{figs:3b}
\end{figure*}

Besides, for the sake of testing the convergence of the solutions (bearing in mind the numerical dissipation effects), the simulation for $Re=17000$ was obtained using seven levels of refinement (shown in Fig.~\ref{figs:3b}).

\begin{figure*}
\centering
\begin{tabular}{cc}
\includegraphics[width=0.4\linewidth]{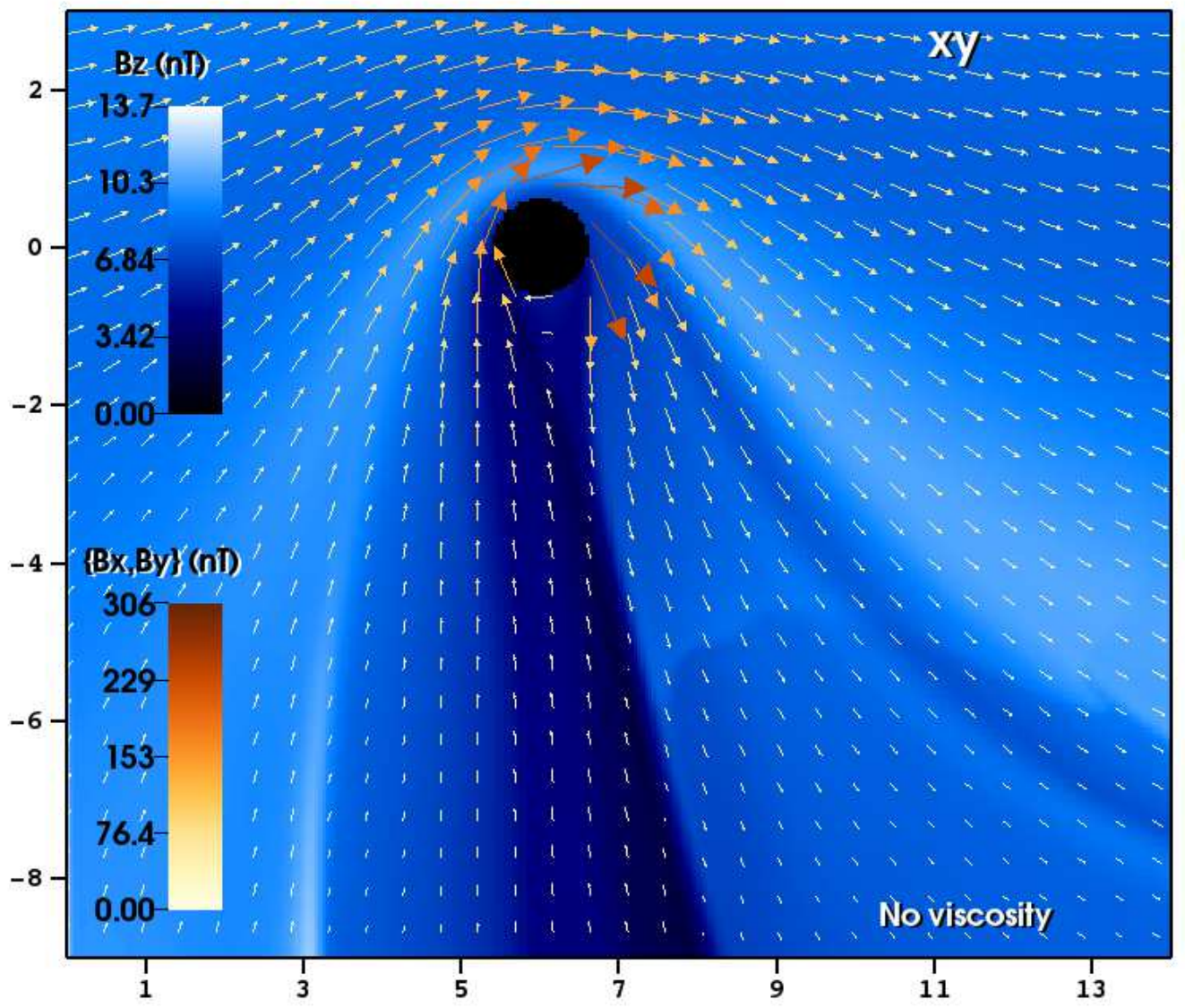}
\includegraphics[width=0.4\linewidth]{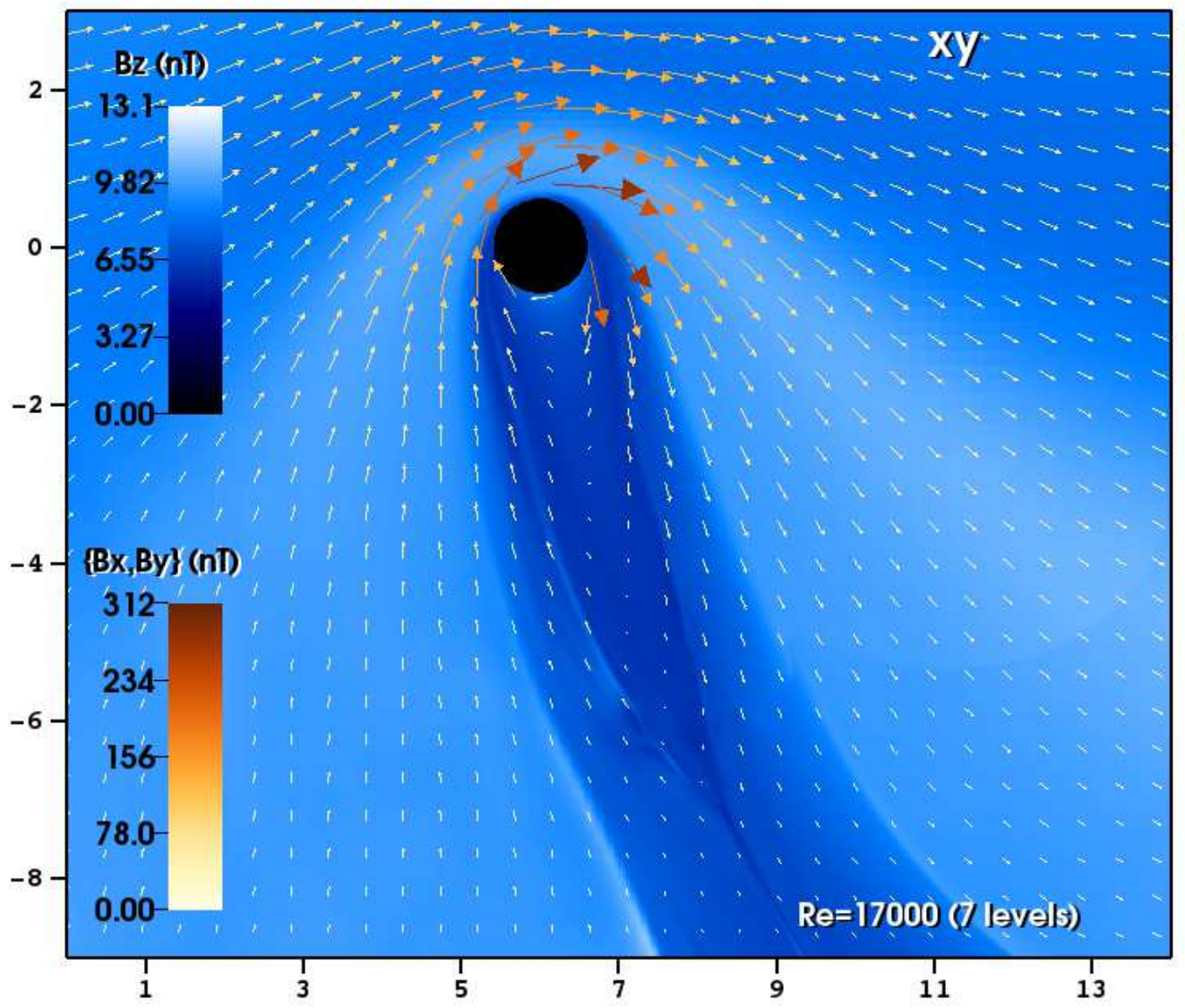}\\
\end{tabular}
\caption{Magnetic field in $\unit{nT}$ at $t=\unit[1400]{s}$ for the cases with no viscosity (left, with five levels) and $\nu=\unit[300]{km^2~s^{-1}}$ (right, with seven levels). The components $B_x$ and $B_y$ are represented as a vector field and $B_z$ in color plot. The dimensions of the box are in $\unit[10^{9}]{cm}$.}
\label{figs:4}
\end{figure*}

Figures~\ref{figs:3} and \ref{figs:3b} show the outlines of the mesh refinement. The most refined areas are along the shocks, as well as around the object and where $\rho$ and $\mathbf{B}$ present variations (see Fig.~\ref{figs:4} too), indicating us that PARAMESH remains stable in MHD simulations with solid objects under the present conditions.

Comparing Figs.~\ref{figs:3} and \ref{figs:3b} we may note that the wake is thinner for seven levels when compared to the other scenarios. Such a behavior is related to the refinement of the solutions: the higher the refinement, the thinner the wakes, with their dimensions converging to a particular value for sufficiently high refinements. With effect, concerning the dimensions of the wake, the simulation with six levels (left panel of Fig.~\ref{figs:8b}) represents an intermediate case between the less and the more refined ones.

\begin{figure}
\centering
\includegraphics[width=0.7\linewidth]{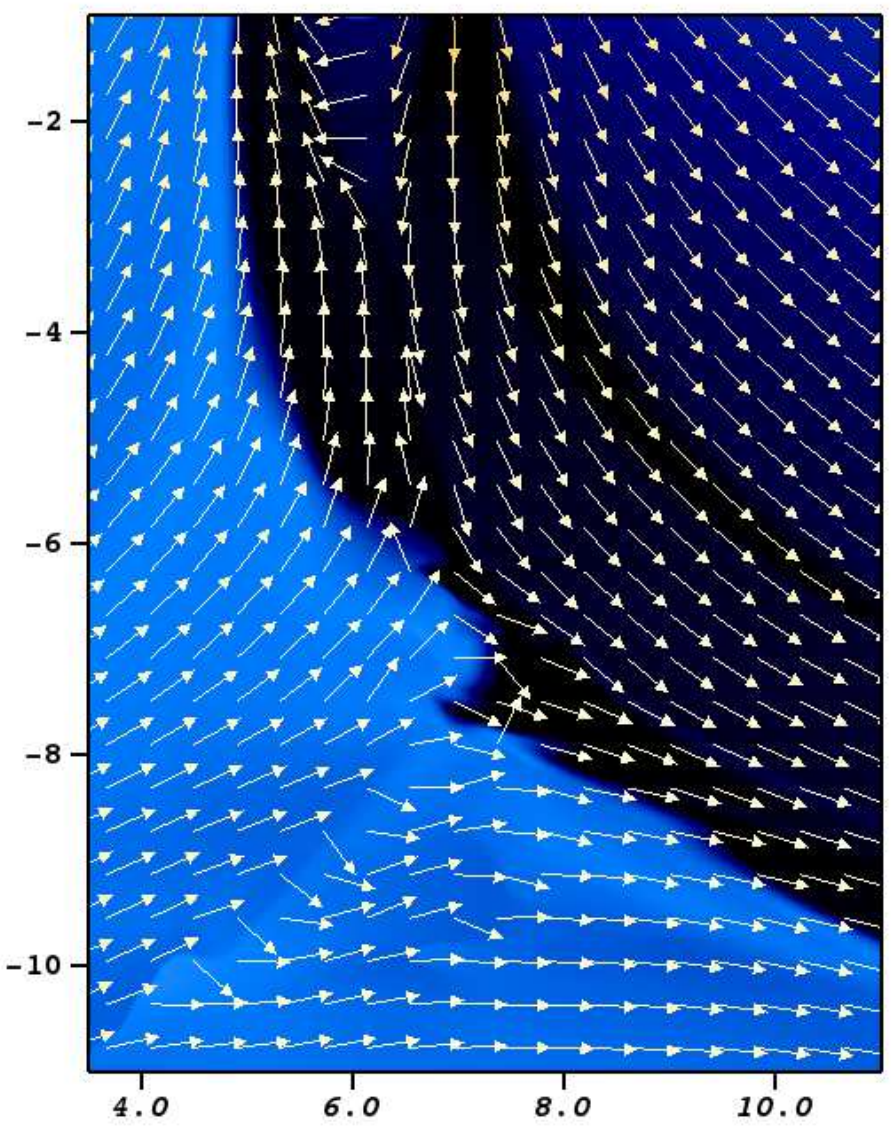}
\caption{Zoom of the structures in the wake of the right bottom panel of Fig.~\ref{figs:3}. The density is shown in colors (same scale as in Fig.~\ref{figs:3}) and the vector field (not scaled by magnitude) represents $\mathbf{B}_{xz}=\{B_{x},B_{z}\}$.}
\label{figs:4b}
\end{figure}

Note the structures in the wake of the right profile of Fig.~\ref{figs:3b}. A closer view of these structures is shown in Fig.~\ref{figs:4b}, which shows the density in colors (same scale as in Fig.~\ref{figs:3} and \ref{figs:3b}) and $\mathbf{B}_{xz}=\{B_{x},B_{z}\}$ as a vector field. The vectors of $\mathbf{B}_{xz}$ are not scaled by magnitude for a better visualization but $|\mathbf{B}_{xz}|$ has a maximum value of $\sim\unit[100]{nT}$ in the region. The behavior of $\mathbf{B}_{xz}$ in Fig.~\ref{figs:4b} is suggestive of a magnetic reconnection process possibly happening in such a region.

It is interesting to observe that, though the stellar wind is parallel to the x-axis, there is no symmetry around $y=0$ in Fig.~\ref{figs:3} and \ref{figs:3b}. We explain this behavior as follows: as the simulation evolves, the plasma starting with velocity $\mathbf{v}=v_{\mbox{\scriptsize{sw}}} \boldsymbol{\hat{\textbf{i}}}$ undergoes magnetic forces due to $B_y$ and $B_z$, causing the emergence of $v_y$ and $v_z$ components in the fluid velocities (though some of $v_y$ and $v_z$ arises from the interaction with the rigid body). Then $B_x$ exerts forces transverse to the x-axis on the portions of the fluid where $v_y$ and $v_z$ are different from zero.

We plot the magnetic field at $t=\unit[1400]{s}$, shown in Fig.~\ref{figs:4}. The perspective is from the xy-plane, with the components $B_x$ and $B_y$ represented as a vector field and $B_z$ in color plot. The left and right panels correspond to the scenarios with no viscosity and with $\nu=\unit[300]{km^2~s^{-1}}$, respectively. In both cases we have $\sqrt{B^2_x+B^2_y}\approx\unit[300]{nT}$ around the planet and $B_z\approx\unit[13]{nT}$ along the bow shock; besides, in the wake $|\mathbf{B}|$ has the lowest values.

Analyzing the initial $\mathbf{B}$ given in Table~\ref{tab:1}, we deduce that the interaction of the wind with the body increases $\sqrt{B^2_x+B^2_y}$ by a factor of $\approx 8.5$, while the values of $B_z$ remains of the same order of magnitude.

\begin{figure*}
\centering
\begin{tabular}{cc}
\includegraphics[width=0.2\linewidth]{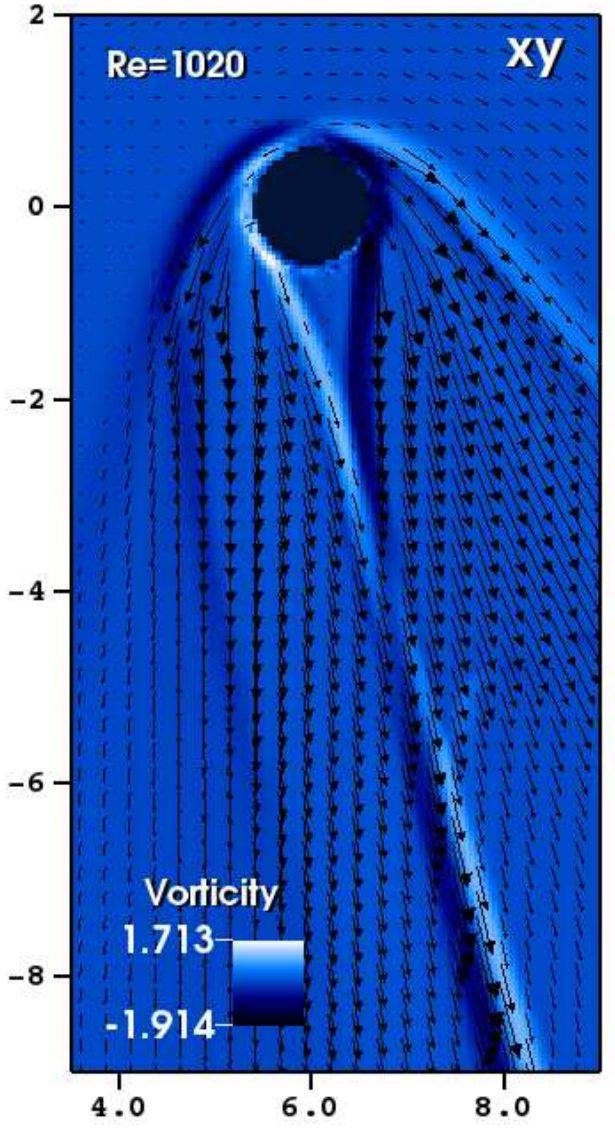}
\includegraphics[width=0.2\linewidth]{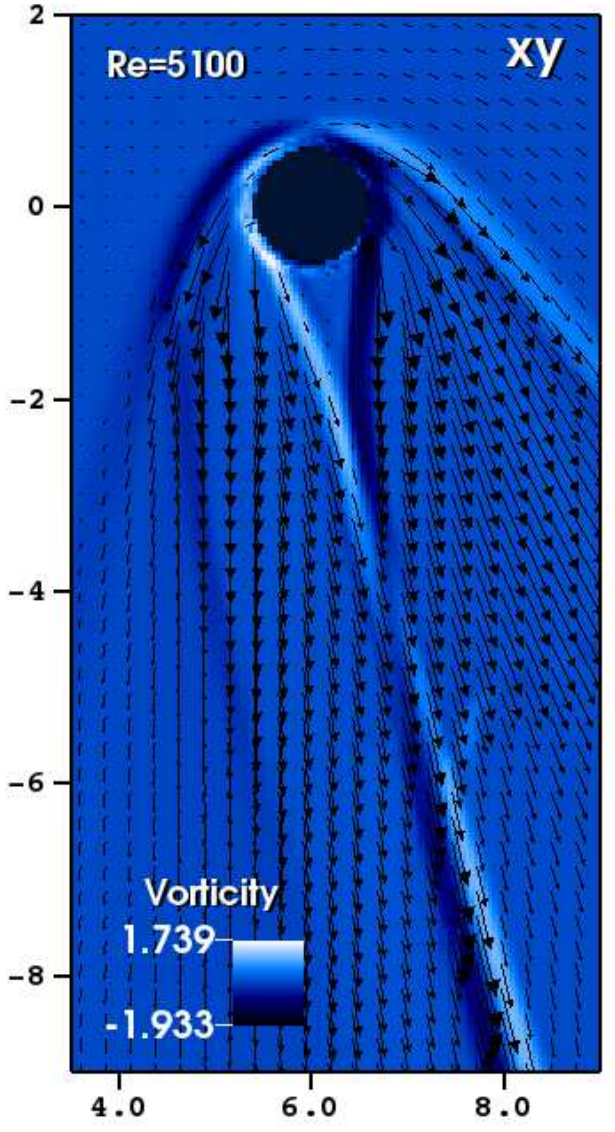}
\includegraphics[width=0.2\linewidth]{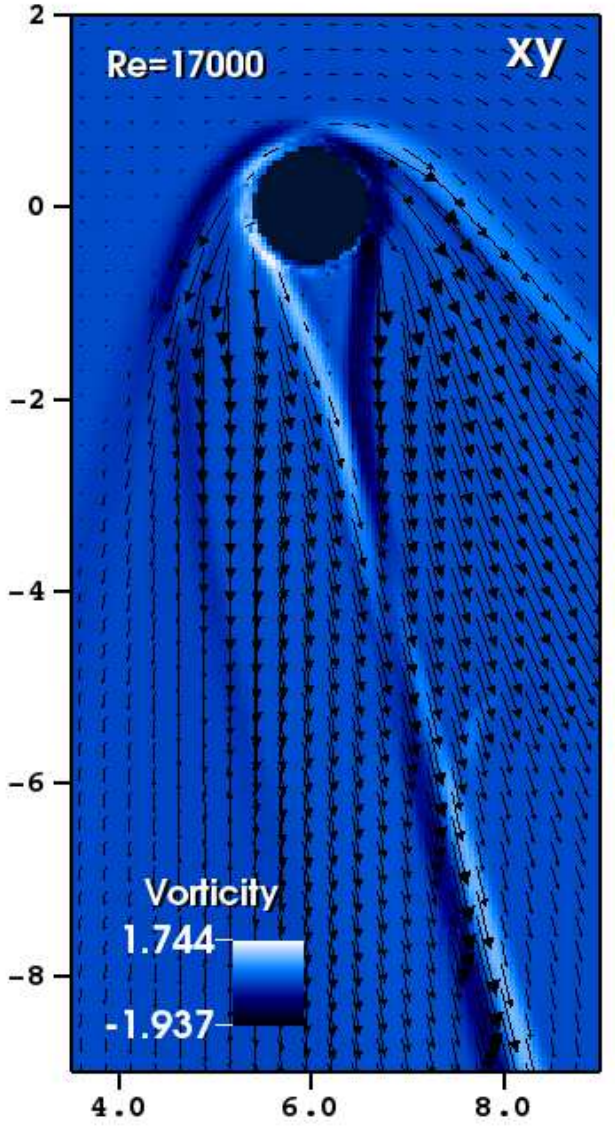}
\includegraphics[width=0.2\linewidth]{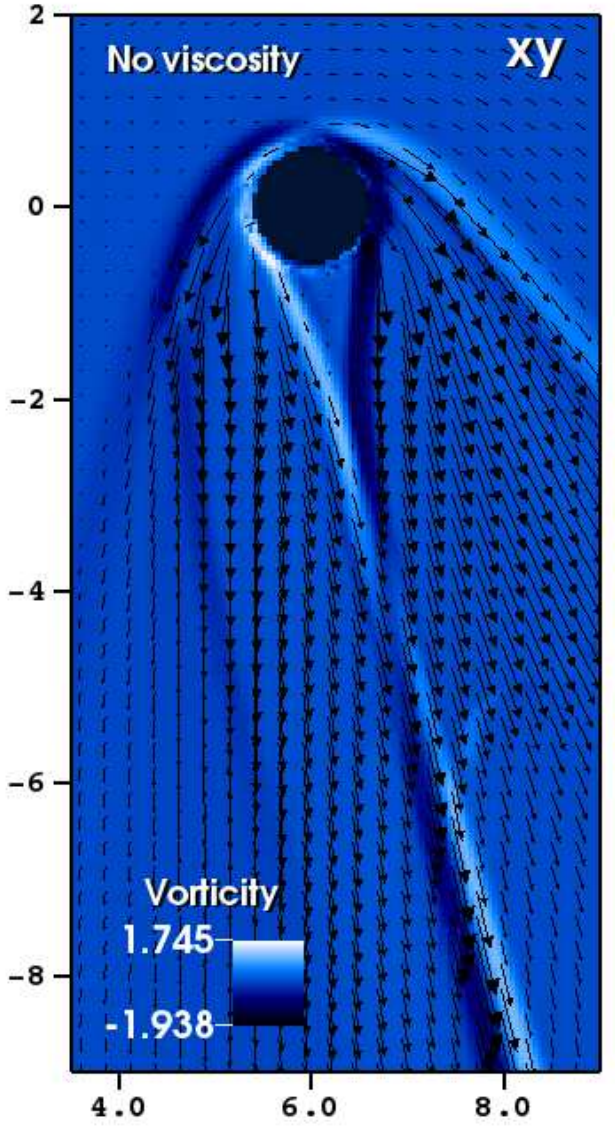}\\
\end{tabular}
\caption{Velocity vector field and vorticity profiles for the MHD simulations in the xy-plane and at $t=\unit[1400]{s}$. From left to right: $\nu=\unit[5000]{km^2~s^{-1}}$ (Re=$1020$), $\nu=\unit[1000]{km^2~s^{-1}}$ (Re=$5100$), $\nu=\unit[300]{km^2~s^{-1}}$ (Re=$17000$) and no viscosity. The maximum $|\mathbf{v}|$ are of $\approx\unit[3.0\times 10^8]{cm~s^{-1}}$ and the dimensions of the box are in $\unit[10^{9}]{cm}$. All the panels were obtained with five levels of refinement.}
\label{figs:5}
\end{figure*}

Figure~\ref{figs:5} shows the velocity vector field and the vorticity profiles for the MHD simulations at $t=\unit[1400]{s}$ in the xy-plane. As in the previous case, four scenarios are considered: with $\nu=\unit[5000]{km^2~s^{-1}}$ (Re=$1020$), $\nu=\unit[1000]{km^2~s^{-1}}$ (Re=$5100$), $\nu=\unit[300]{km^2~s^{-1}}$ (Re=$17000$) and with no viscosity. The four scenarios were generated with five levels of refinement. The maximum value of the velocity in the four scenarios of Fig.~\ref{figs:5} are of $\approx\unit[3.0\times 10^8]{cm~s^{-1}}$. Here we have $L\approx\unit[1.7\times 10^{9}]{cm}$ for the four scenarios; the maximum $|\omega_z|$ slightly increases with $Re$ and its values may be observed in Fig.~\ref{figs:5} and Fig.~\ref{figs:9}.

\subsection{MHD scenario with initial $B_x=0$}
\label{BxEqualZero}

\begin{figure*}
\centering
\begin{tabular}{cc}
\includegraphics[width=0.44\linewidth]{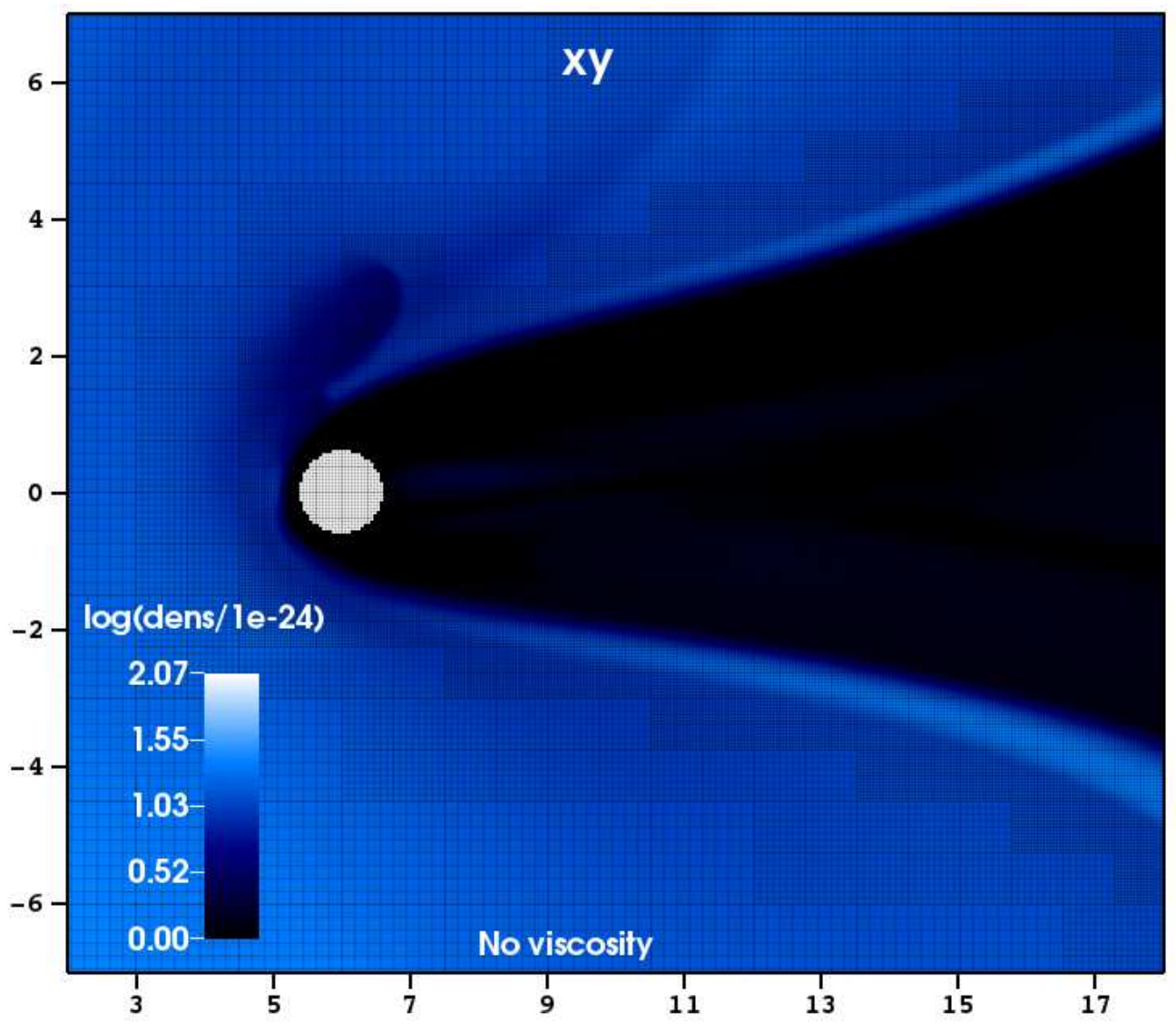}
\includegraphics[width=0.44\linewidth]{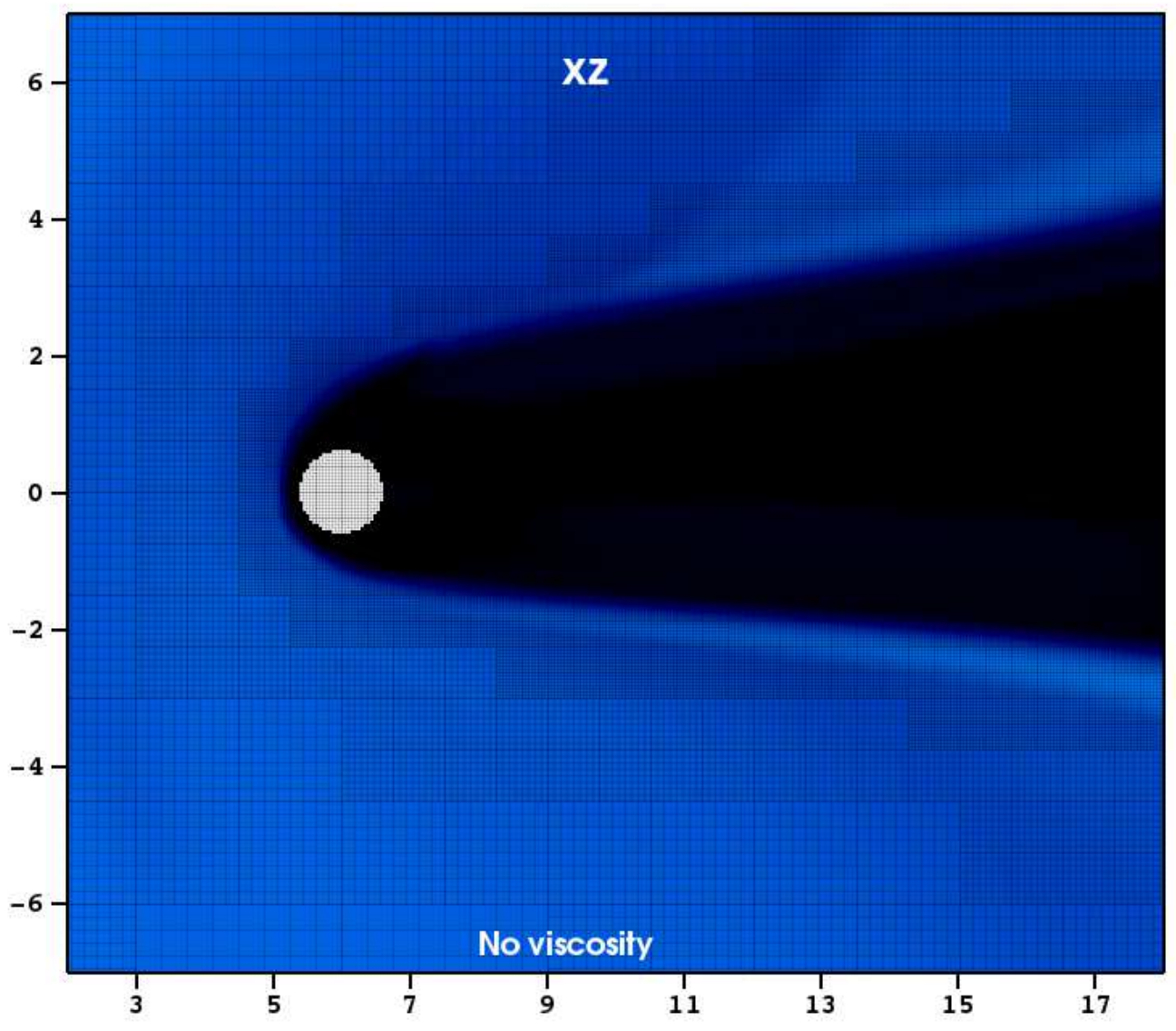}\\
\includegraphics[width=0.44\linewidth]{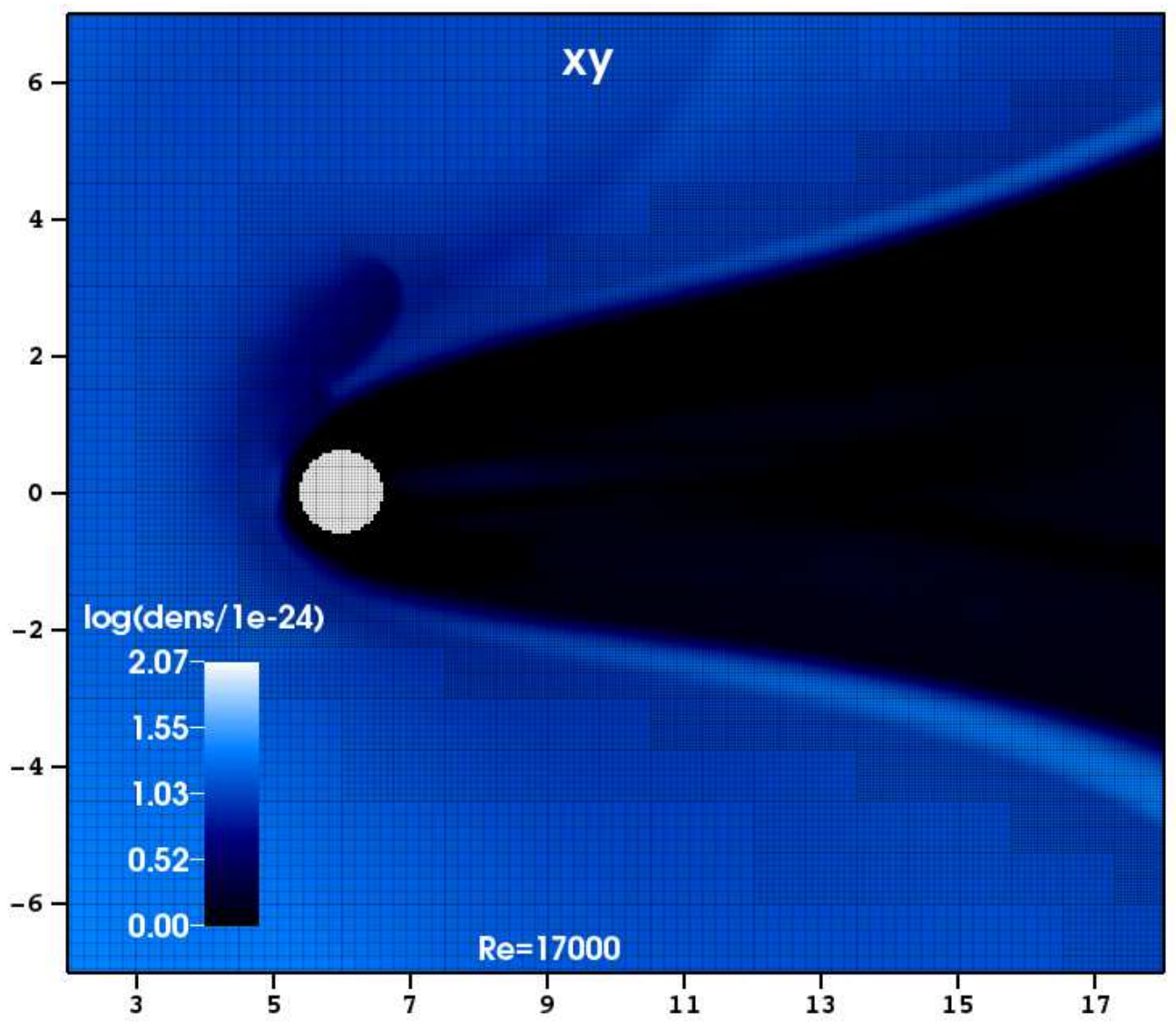}
\includegraphics[width=0.44\linewidth]{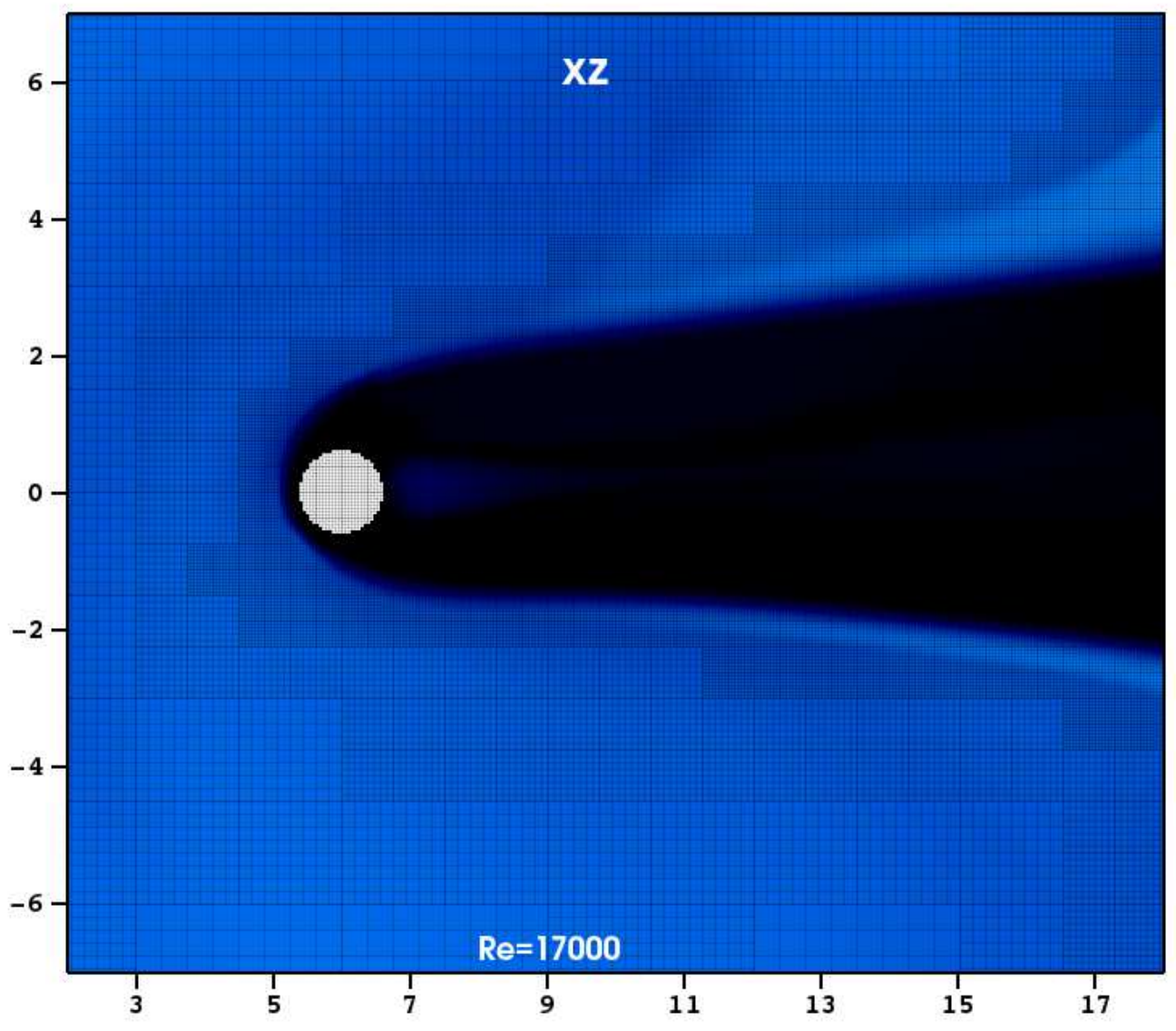}\\
\end{tabular}
\caption{Same as in Fig.~\ref{figs:3} but considering an initial $B_x=0$ and with five levels of refinement in both scenarios.}
\label{figs:6}
\end{figure*}

As an extra result, we performed simulations using the same parameters as the ones shown in Subsection~\ref{BxDiffZero} but considering $B_x=0$ in the initial conditions. Though this scenario is not realistic, once from Parker's model $B_r/B_{\phi} \ll 1$ only for large heliocentric distances, it will help us to observe the influence of the transversal components of $\mathbf{B}$ on the interaction of the wind with the planet. The densities and mesh refinement with five levels at $t=\unit[1400]{s}$ are shown in Fig.~\ref{figs:6}. We note that there is symmetry about $y=0$ and, as in the previous MHD case, it is formed a discernible bow shock. The bow shocks in Fig.~\ref{figs:6} are characterized by $\rho=\unit[1\times 10^{-23}]{g~cm^{-3}}$ and $p=\unit[2.0\times 10^{-8}]{dyn~cm^{-2}}$, while the wakes have $\rho\sim \unit[10^{-24}]{g~cm^{-3}}$ and $p=\unit[7.0\times 10^{-8}]{dyn~cm^{-2}}$ at their central regions.

Concerning the mesh refinement, we see that the most refined areas are around the shocks, object and wakes, following the variations of $\rho$ and $\mathbf{B}$. As in the previous MHD scenario, we point out the stability of the numerical schemes that are integrated in 
the PARAMESH structure in this case.

\begin{figure*}
\centering
\begin{tabular}{cc}
\includegraphics[width=0.4\linewidth]{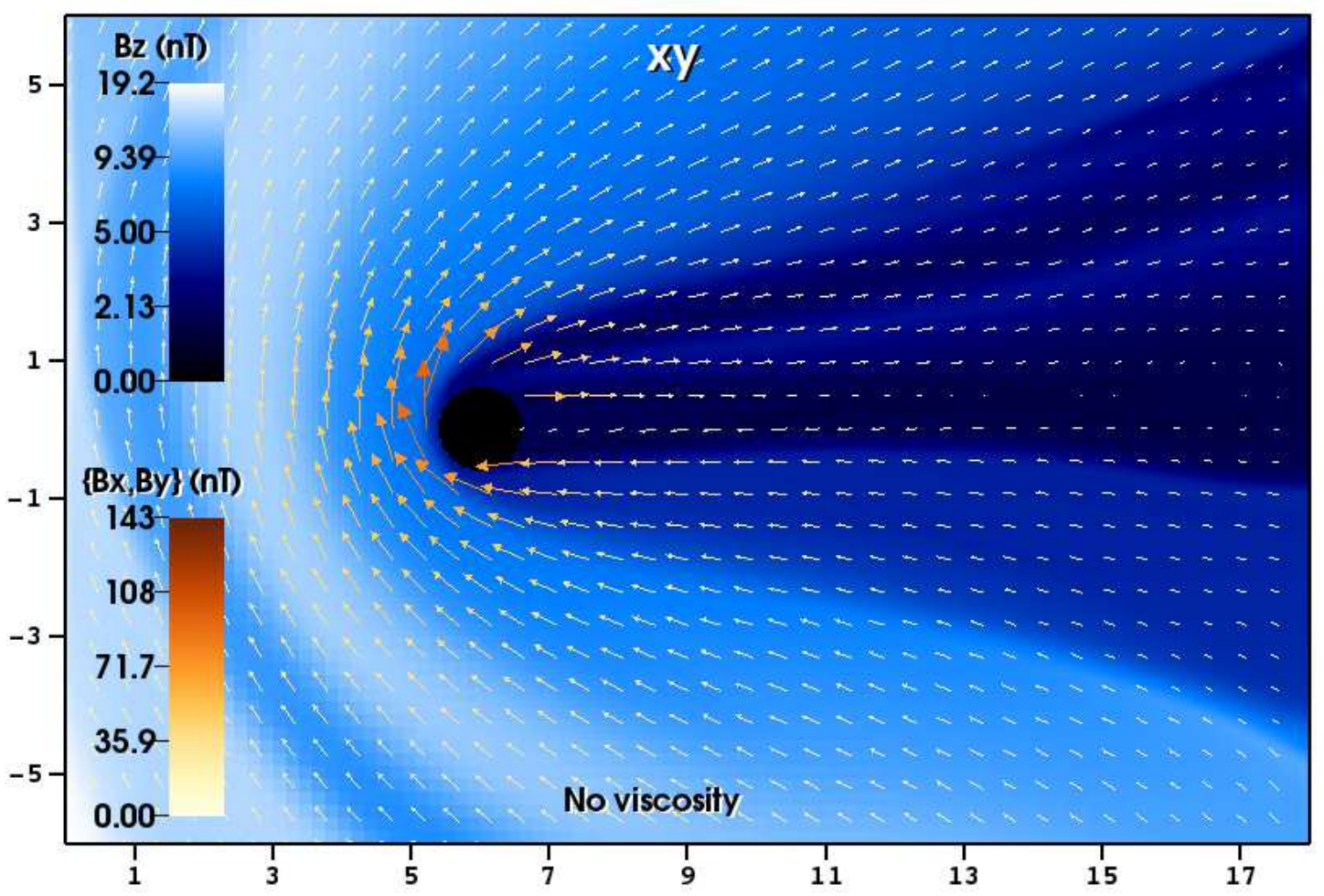}
\includegraphics[width=0.4\linewidth]{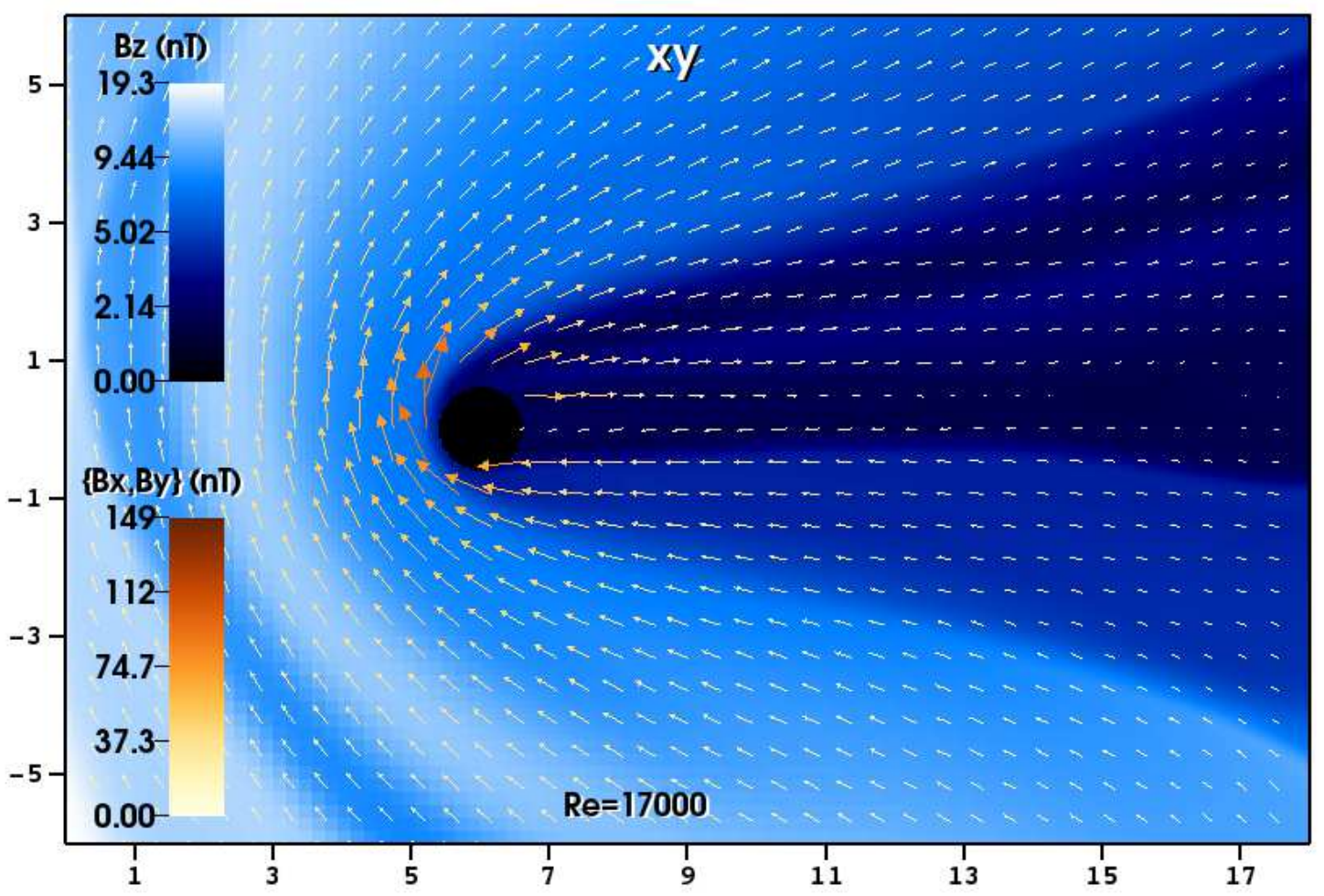}\\
\end{tabular}
\caption{Same as in Fig.~\ref{figs:4} but considering an initial $B_x=0$ and with five levels of refinement in both scenarios.}
\label{figs:7}
\end{figure*}

Figure~\ref{figs:7} for $\mathbf{B}$ follows the same scheme of Fig.~\ref{figs:4}. We have $\sqrt{B^2_x+B^2_y}=\unit[143]{nT}$ (left panel) and $\sqrt{B^2_x+B^2_y}=\unit[149]{nT}$ (right panel) around the planet and $|B_z|$ has maximun values of $\approx\unit[19]{nT}$. We see that $\sqrt{B^2_x+B^2_y}$ is increased by the factors $\approx 4.1$ (left panel) and $\approx 4.3$ (right panel) when compared to its initial value; $B_z$ reachs values which are $1.9$ higher than the initial one. In Figs.~\ref{figs:4} and \ref{figs:7} we observe a pattern of circulation of $\mathbf{B}$ in the xy-plane. Particularly, in the inner regions of the wake we have magnetic field lines which are oppositely directed and are close to each other. Under certain circumstances, such a behavior could potentially create suitable conditions for the onset of magnetic reconnection.

\begin{figure*}
\centering
\begin{tabular}{cc}
\includegraphics[width=0.4\linewidth]{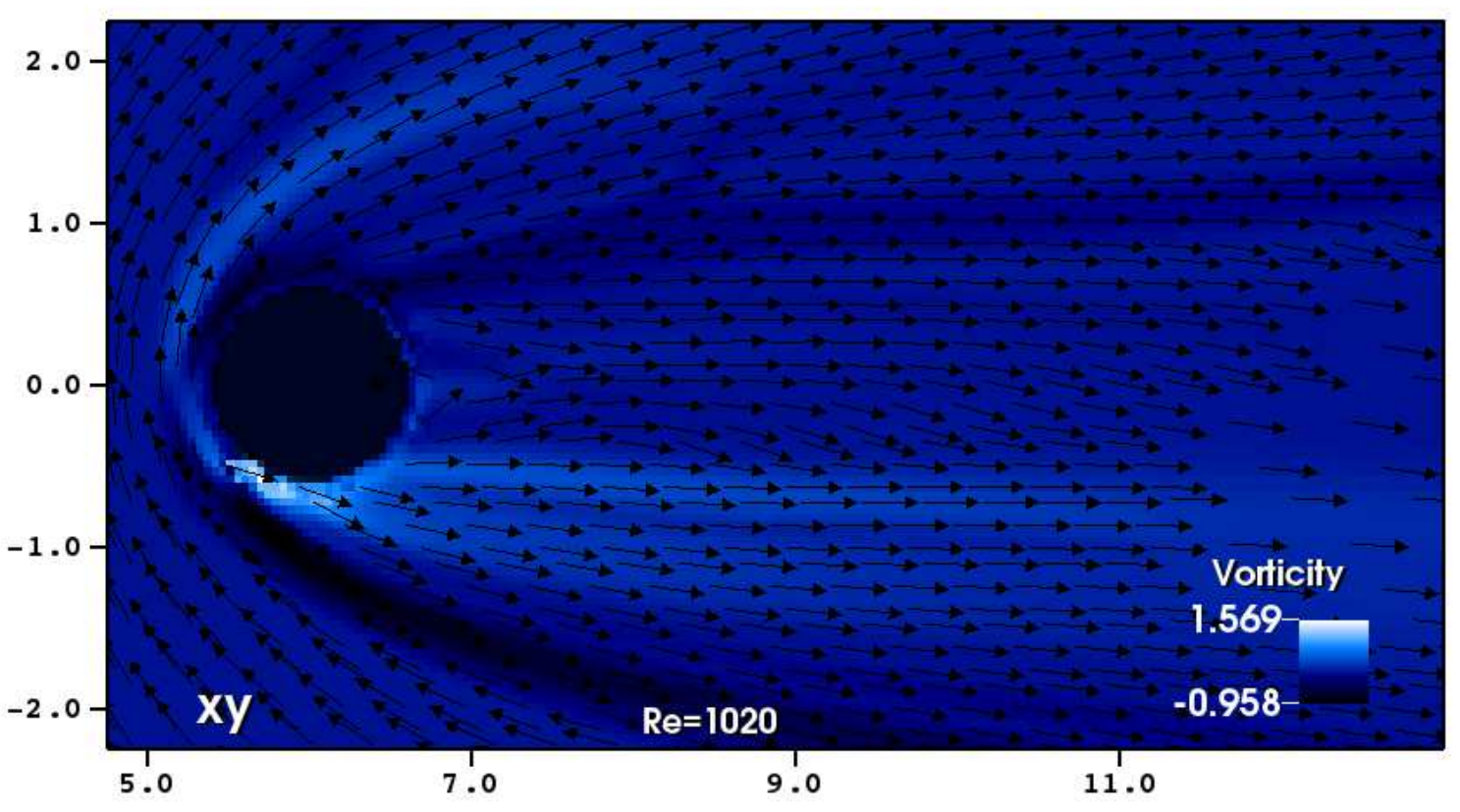}
\includegraphics[width=0.4\linewidth]{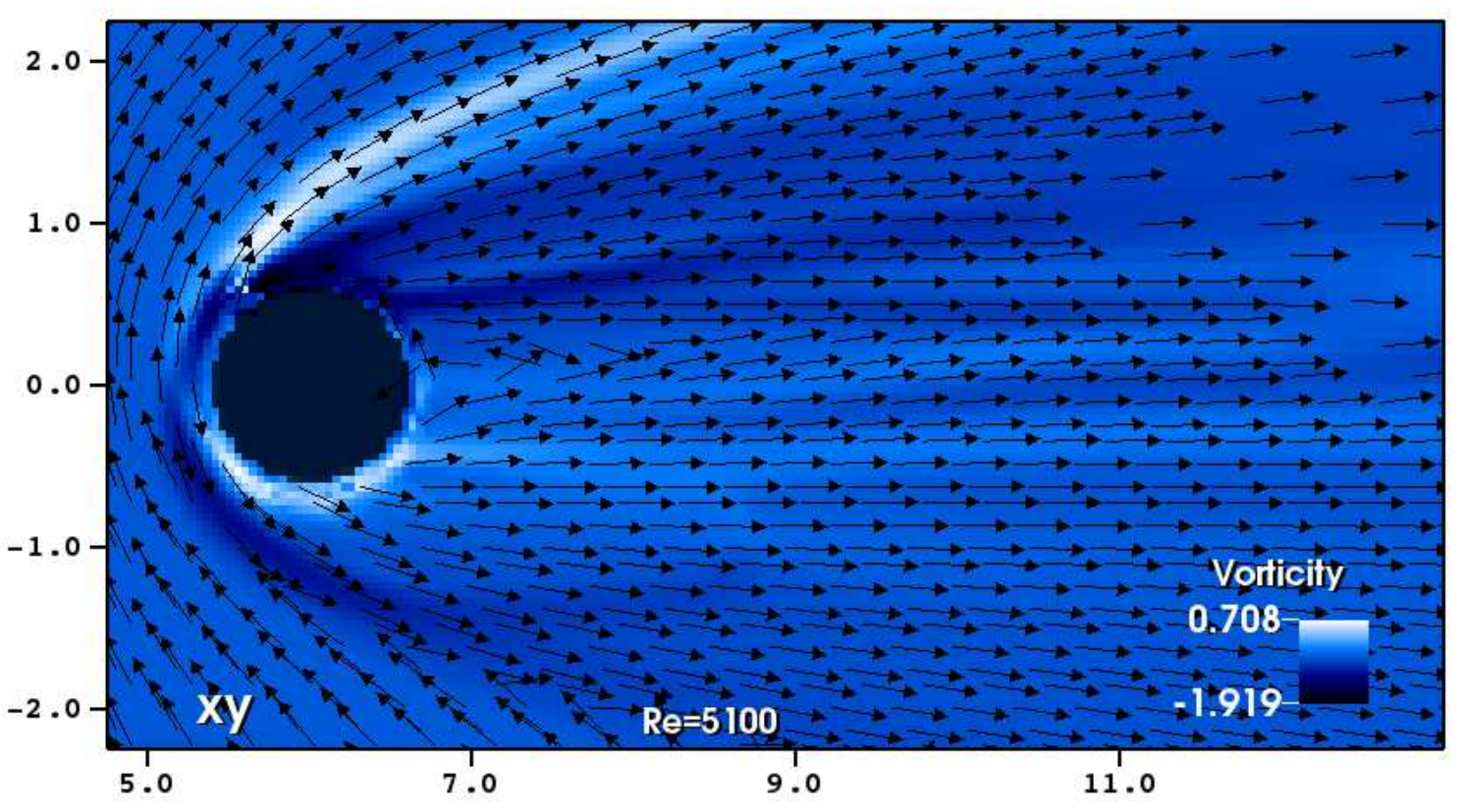}\\
\includegraphics[width=0.4\linewidth]{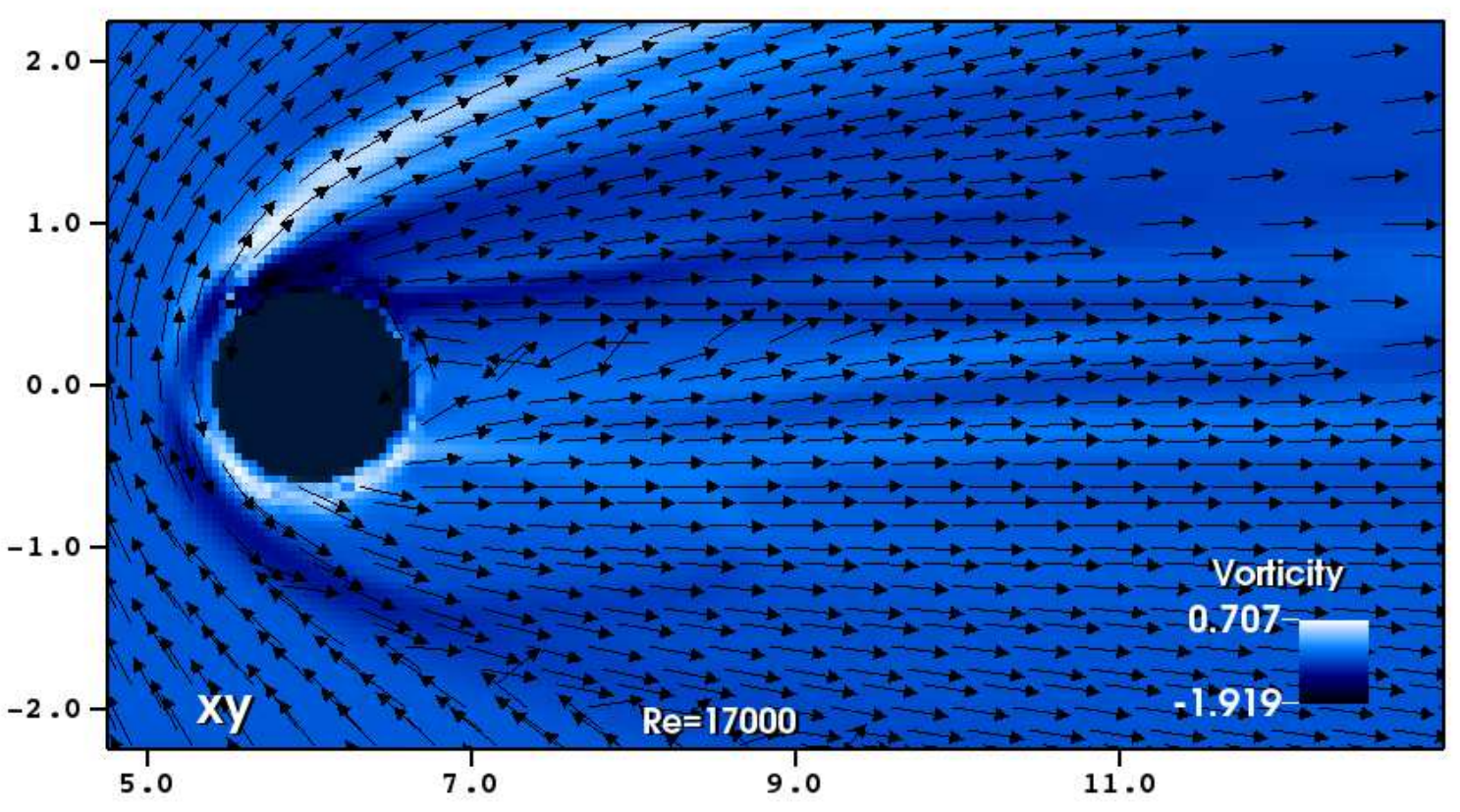}
\includegraphics[width=0.4\linewidth]{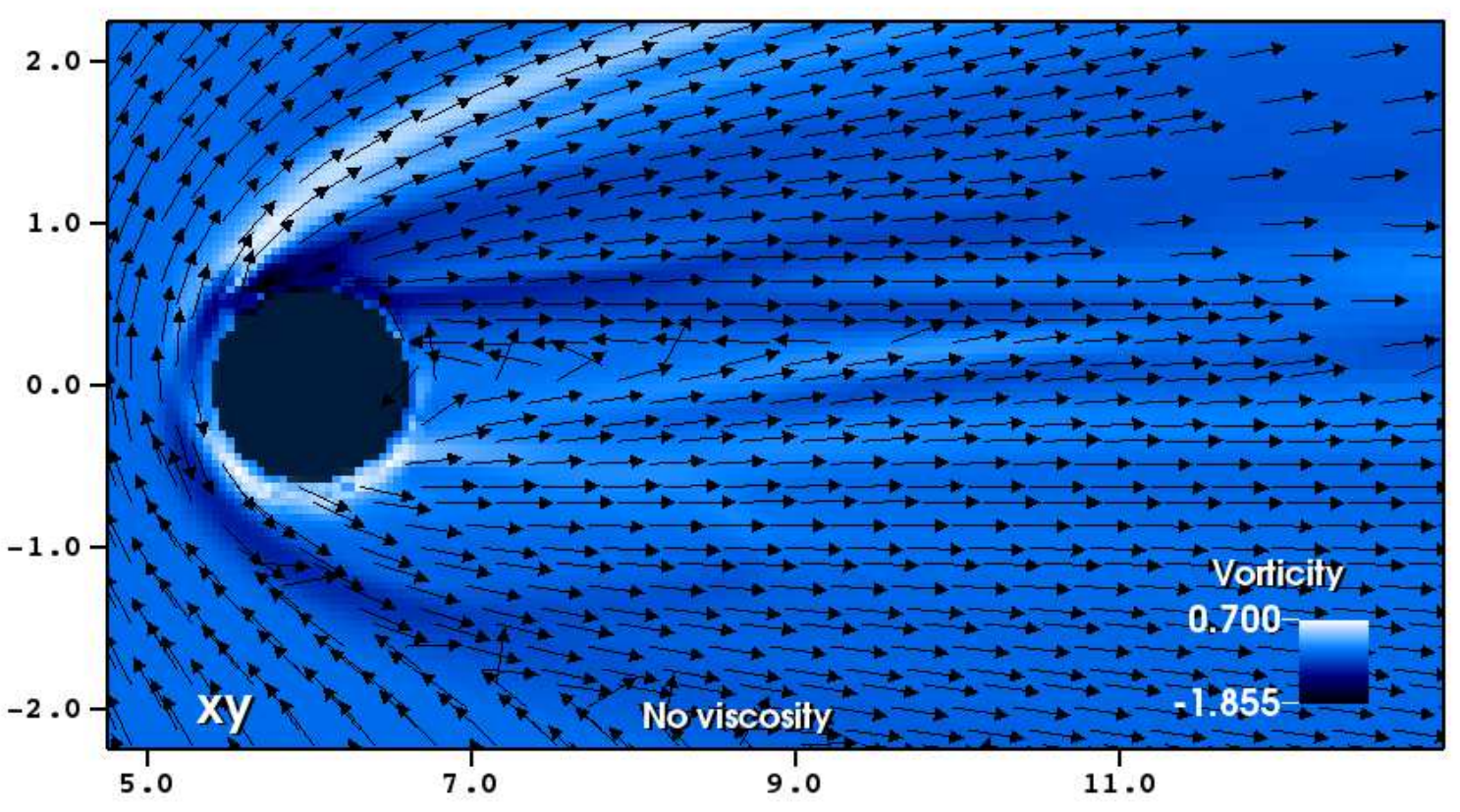}\\
\end{tabular}
\caption{Same as in Fig.~\ref{figs:5} but considering an initial $B_x=0$.}
\label{figs:8}
\end{figure*}

The velocity vector field and the vorticity profiles at $t=\unit[1400]{s}$ are given in Fig.~\ref{figs:8}. The scheme is similar to Fig.~\ref{figs:5}. However, note that here, in order to better observe the recirculation zones, the arrows of the velocity fields are not scaled by magnitude. We have $L=\unit[5.8\times 10^{8}]{cm}$, $L=\unit[9.6\times 10^{8}]{cm}$ and $L=\unit[1.7\times 10^{9}]{cm}$ for increasing values of $Re$ ($L=\unit[3.3\times 10^{9}]{cm}$ with no viscosity); $|\omega_z|$ increases between $Re=1020$ and $Re=5100$ and its values are shown in Fig.~\ref{figs:8} and Fig.~\ref{figs:9}.

In the MHD simulations there is the formation of a low-density layer between the object and the interacting stellar wind, which has a minimum thickness of, for example, $\approx \unit[1.3\times 10^{8}]{cm}$ in the upper panels of Fig.~\ref{figs:3} and $\approx \unit[2.0\times 10^{8}]{cm}$ in Fig.~\ref{figs:6}. As this phenomenon is not present in the hydrodynamic case we deduce that it is mainly due to the action of $\mathbf{B}$ which, moreover, reach its maximum values in the areas adjacent to the object.

In \cite{spreiter:1970}, the scenario of the interaction of the solar wind with Venus (considered as having no significant magnetic field) presents a bow shock similar to the ones in our MHD simulations; besides, in such a scenario there is a low-density layer of thickness $\unit[5\times 10^7]{km}$ between the shock and Venus. According to the authors, that layer is formed when the ionosphere of the planet deflects the solar wind, preventing it to hit the surface. Though in our model the planet has no atmosphere, the action of $|\mathbf{B}|$ around the body produced a similar effect, as explained in the previous paragraph.

The influence of the viscosity on the length of the recirculation zone in MHD simulations may be found, for example, in \cite{grigoriadis:2010}. In this paper, the MHD scenarios (with streamwise and transverse magnetic fields) for $Re=100$ have, generally speaking, higher $L$ when compared to the cases where $Re=40$.

\subsection{Influence of the boundaries on the simulations}
\label{comparison}

\begin{figure*}
\centering
\begin{tabular}{ccc}
\includegraphics[width=0.32\linewidth]{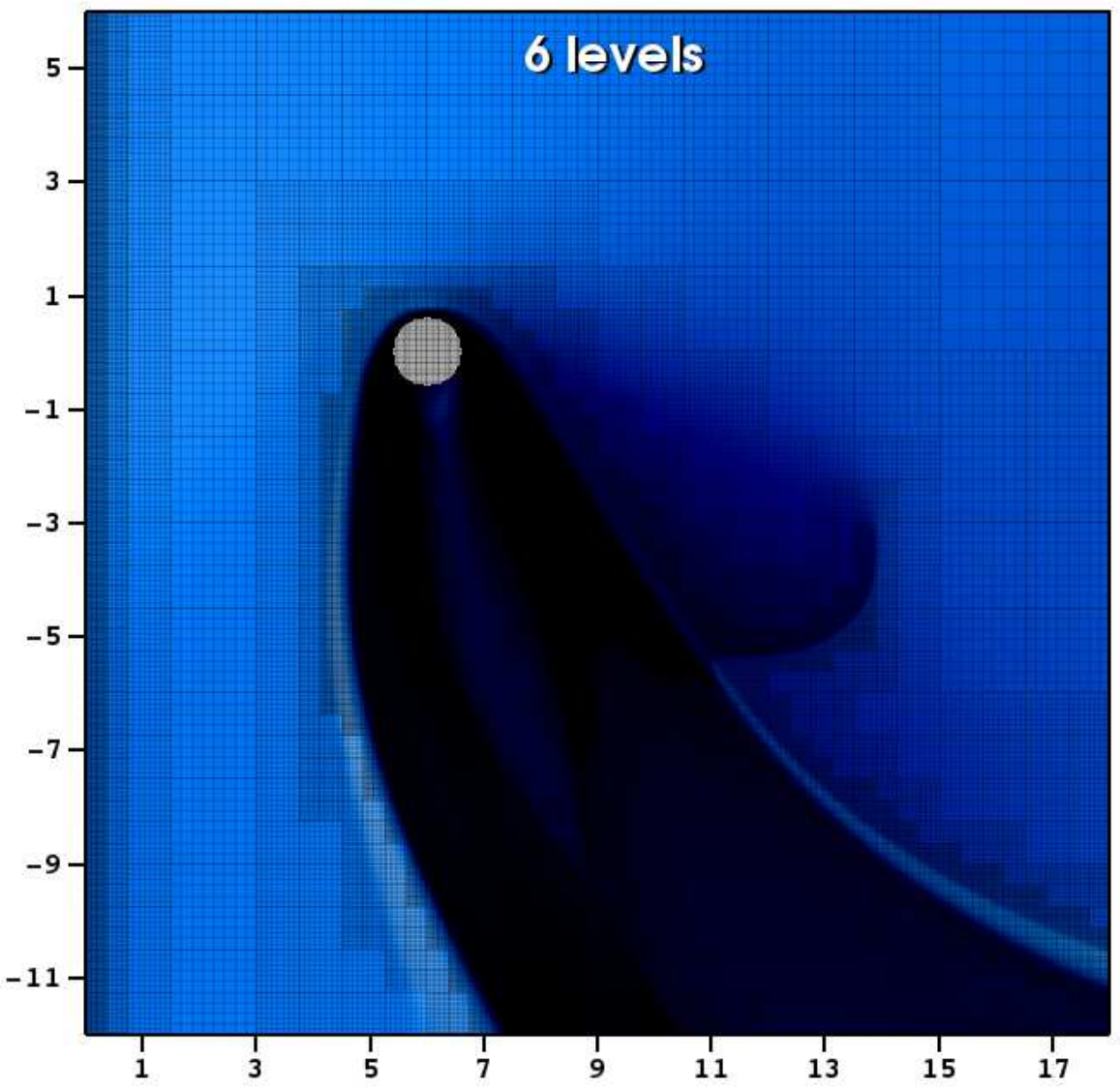}
\includegraphics[width=0.32\linewidth]{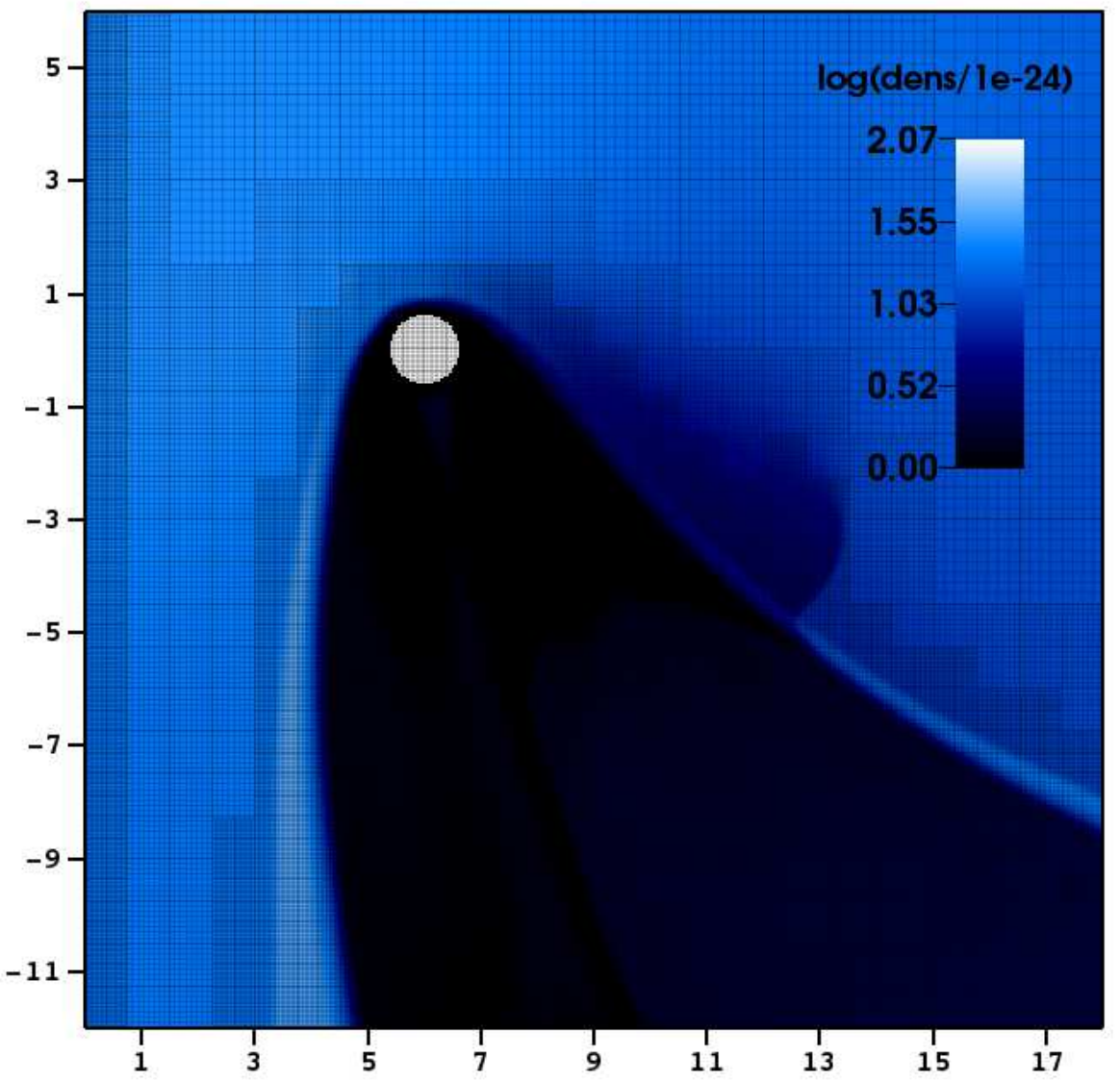}
\includegraphics[width=0.32\linewidth]{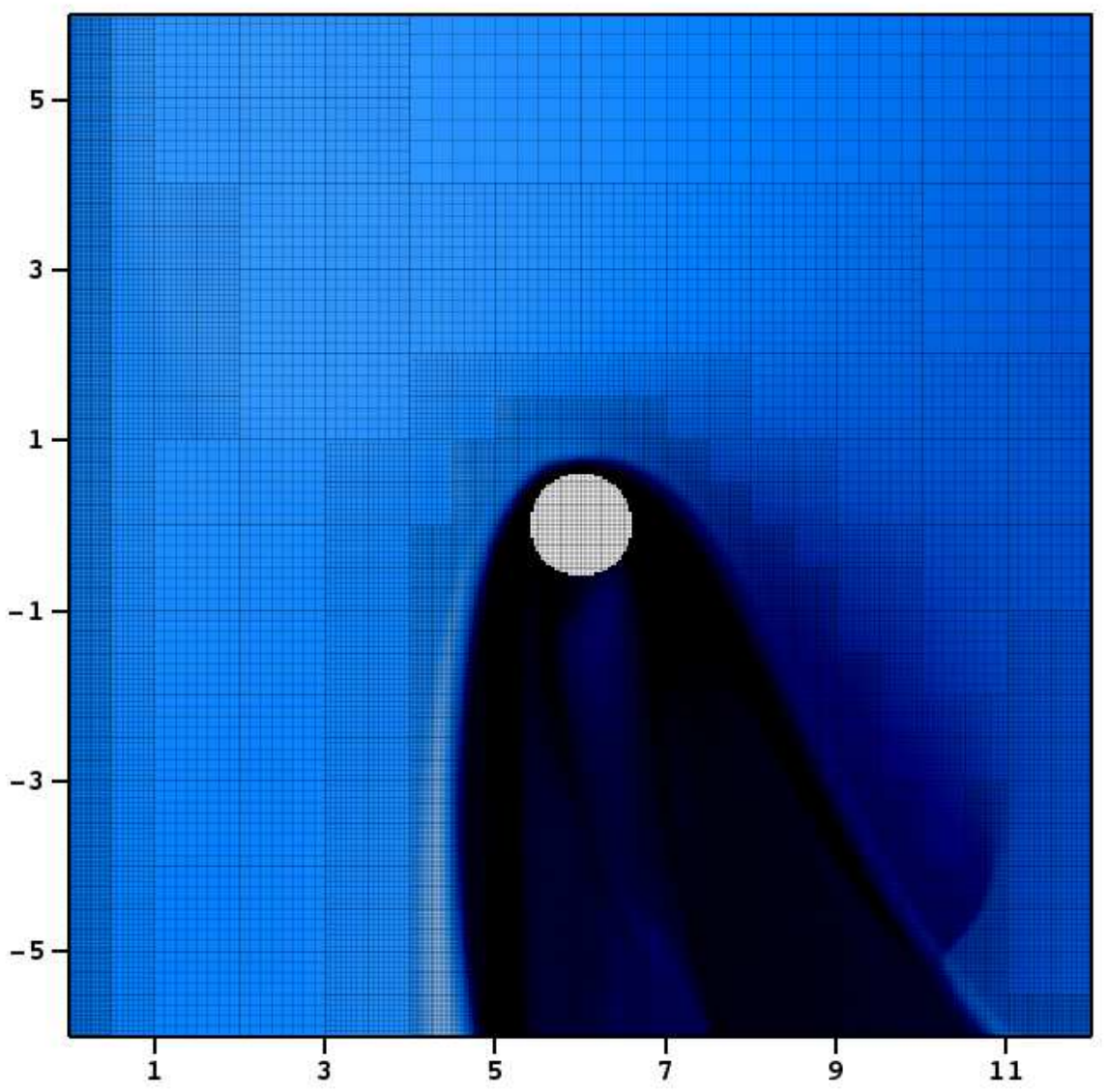}\\
\end{tabular}
\caption{Center and right: simulations for the MHD case with initial $B_x=\unit[32.5]{nT}$, $\nu=\unit[300]{km^2~s^{-1}}$ and five levels of refinement using domains with two sizes; left: the same domain as in the center panel but using six levels. The scale of densities are the same in the three panels and the profiles were taken at $\unit[1400]{s}$.}
\label{figs:8b}
\end{figure*}

Though we are using outflow boundary conditions, it would in principle be possible that some interaction at the borders could propagate back to the domain and influence the results of the simulations. In order to investigate the influence of the boundaries on our results, we performed the simulation of the MHD scenario with initial $B_x=\unit[32.5]{nT}$, $\nu=\unit[300]{km^2~s^{-1}}$ and five levels of refinement using domains with two sizes: $x\in [0.0,18.0] \unit[\times 10^{9}]{cm}$, $y\in [-12.0,6.0] \unit[\times 10^{9}]{cm}$, $z\in [-12.0,6.0] \unit[\times 10^{9}]{cm}$ and $x\in [0.0,12.0] \unit[\times 10^{9}]{cm}$, $y\in [-6.0,6.0] \unit[\times 10^{9}]{cm}$, $z\in [-6.0,6.0] \unit[\times 10^{9}]{cm}$. Figure \ref{figs:8b} presents the density panels at $t=\unit[1400]{s}$ for the bigger (center) and smaller (right) domains. Besides, for the sake of comparison, we show a simulation with the same domain and conditions than the one of the center panel but using six levels of refinement (left.) Note that the center profile is the same as the one presented in Fig.~\ref{figs:3} (bottom left panel). It was shown here again to facilitate a visual comparison.

From Fig.~\ref{figs:8b} we note that the center and right profiles have essentially the same characteristics; we do not observe patterns which would potentially be caused by ``back reactions'' of the boundaries. The patterns in the form of shocks in the right bottom of the panels are created near the planet at the first instants of the simulation and propagate from the left. In Fig.~\ref{figs:6} the reader may observe similar patterns above the planet in the left panels.

Though the center and right panels in Fig.~\ref{figs:8b} have similar characteristics, we may note that they are not equal. We explain the difference between that two cases as follows: in both scenarios we started with the same number of blocks, that is, $3\times3\times3$ in the first level and reaching to $48\times 48\times 48$ in the fifth (see Section~\ref{numerical}). So, the domain at right in Fig.~\ref{figs:8b} is smaller than the other but has the same number of blocks, such that it seems ``more refined'' (see the discussion in Subsection~\ref{BxDiffZero}). With effect, we may compare the center and right profiles in Fig.~\ref{figs:8b} to the left one and to the cases with seven levels of refinement in Fig.~\ref{figs:3}. Particularly, note the similarity between the right and left profiles in Fig.~\ref{figs:8b}. We conclude that the size of the domain do has influence on the results in the sense of refinement, as explained above.


\section{Conclusions}
\label{conclusions}

In this paper we simulated the interaction between the wind produced by a sun-like star and a non-magnetized planet. Such a planet has the approximate size of Earth and an orbital radius of $\unit[0.39]{AU}$, which corresponds to the mean distance between the sun and Mercury. We used the FLASH code to simulate hydrodynamic and MHD scenarios, having as purpose to implement and test the inclusion of a solid and stationary object in MHD simulations in this code. The results presented here are new and interesting once the tool for simulating solid bodies in FLASH is currently implemented and tested for hydrodynamic cases only.

\begin{figure*}
\centering
\begin{tabular}{cc}
\includegraphics[width=0.47\linewidth]{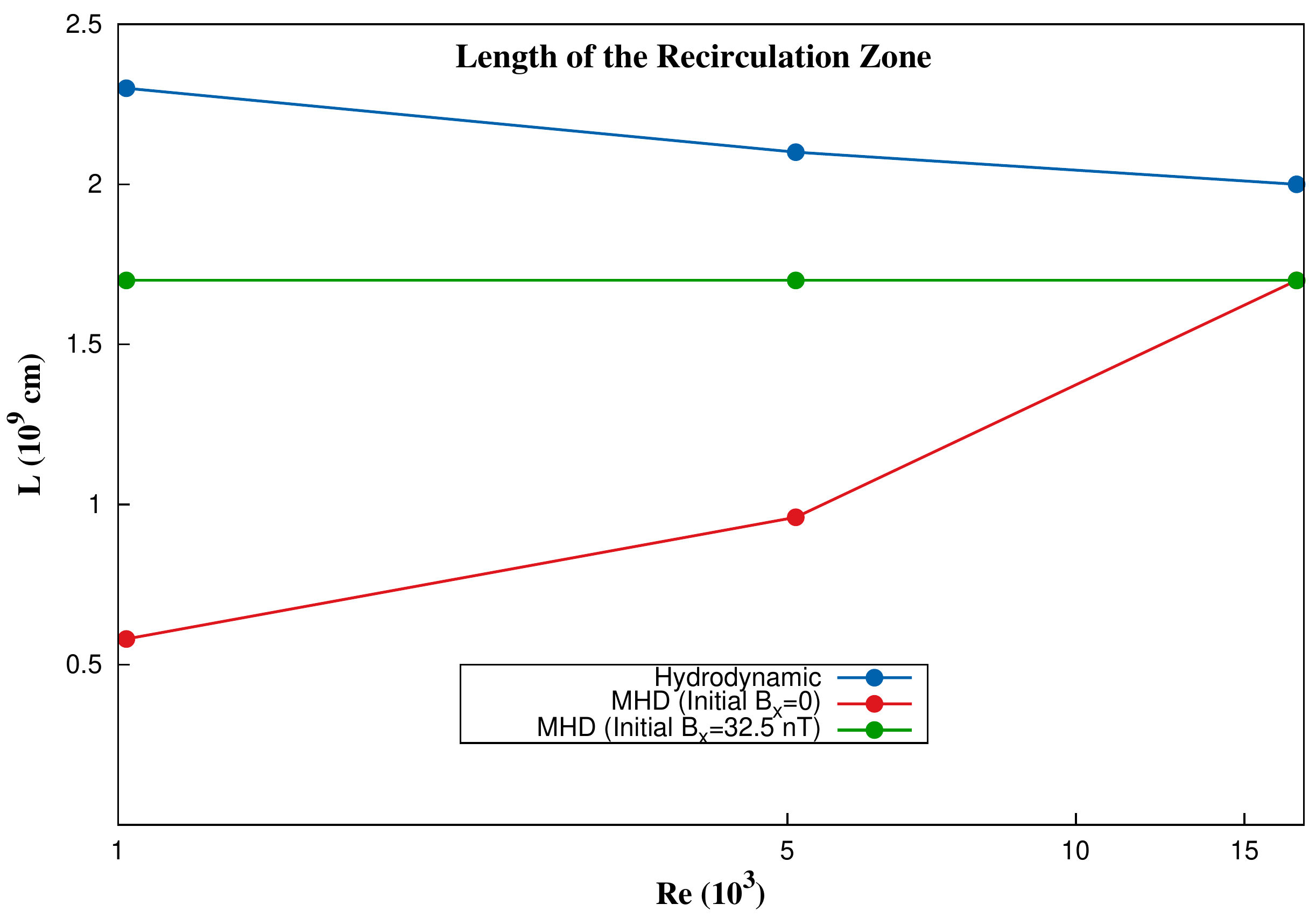}
\includegraphics[width=0.47\linewidth]{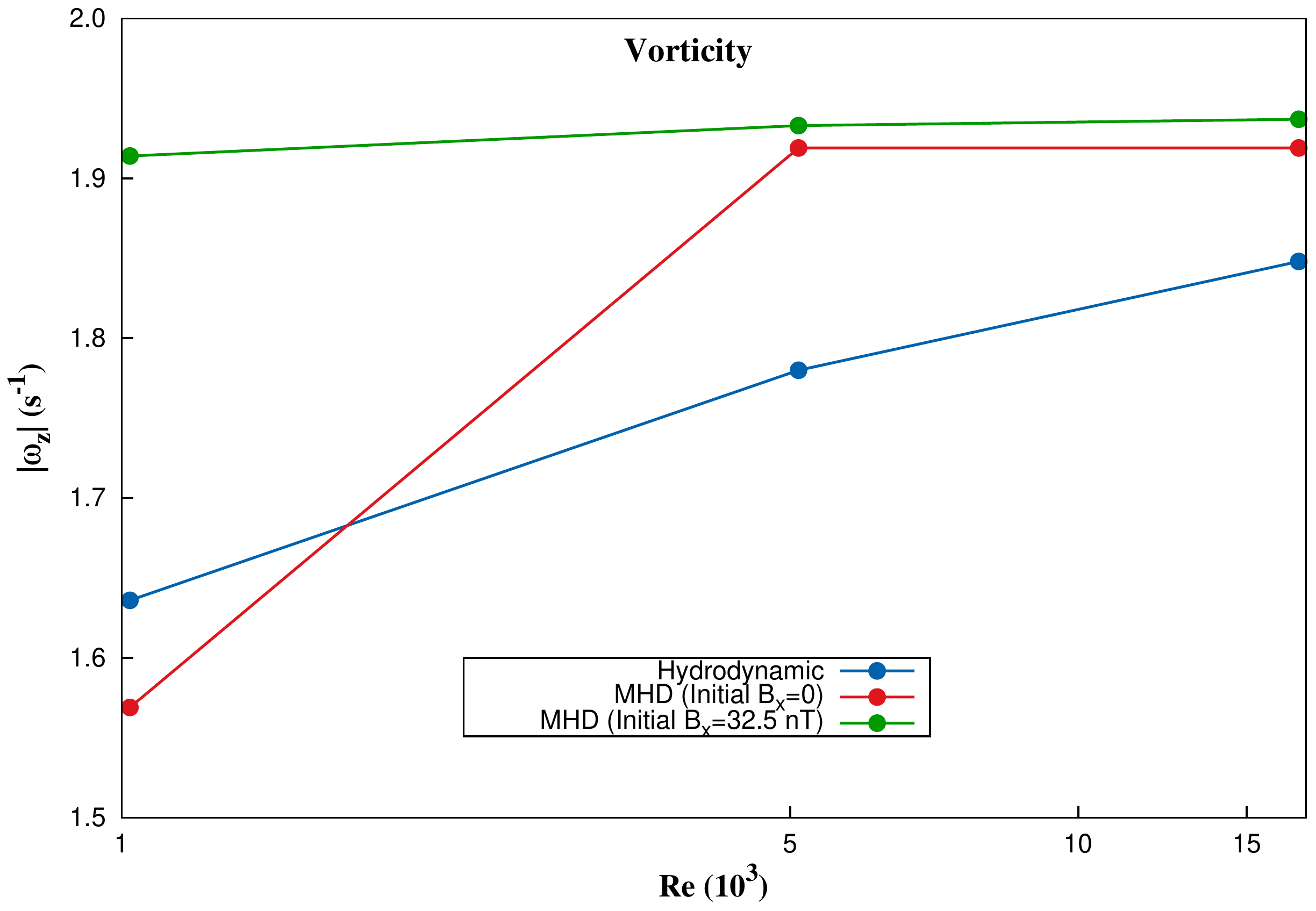}\\
\end{tabular}
\caption{Left: length of the recirculation zone vs. Reynolds number; right: vorticity vs. Reynolds number. For a better visualization, $|\omega_z|$ for the hydrodynamic case was multiplied by four.}
\label{figs:9}
\end{figure*}

The hydrodynamic simulation used as initial parameters of the domain the values of $\mathbf{v}$, $\rho$ and $p$ shown in Table~\ref{tab:1}, besides a maximum time of $\unit[1200]{s}$ and five levels of refinement. We presented the profiles of $\rho$ in the xy-plane for the scenarios with no viscosity and with $\nu=\unit[300]{km^2~s^{-1}}$; besides, we plotted the velocity fields and the vorticity for $Re=1020$, $Re=5100$, $Re=17000$ and no viscous. The differences in the profiles of $\rho$ between the two cases are very small, while the velocity fields indicated us that $L$ slightly decreases (and $|\omega_z|$ increases) as $Re$ increases. See Fig.~\ref{figs:9}.

For the MHD scenario we considered the same initial parameters of the domain as in the hydrodynamic simulation besides adding $\mathbf{B}$. We used Parker's model to define the initial $B_x$ and $B_y$, whereas the values of $B_z$ was estimated by means of observations from the spacecrafts MESSENGER and Helios\cite{korth:2011}, giving $B_x=\unit[32.5]{nT}$, $B_y=\unit[12.7]{nT}$ and $B_z=\unit[10]{nT}$. As an extra MHD result, we simulated the case where we have initially $B_x=0$ which, though is not realistic, helped us to investigate the influence of the transversal components of $\mathbf{B}$ on the simulations.

For the MHD simulations with initial $B_x=\unit[32.5]{nT}$ we shown the profiles of $\rho$ and the outlines of the mesh refinement in the xy and xz-plane: using five levels for the cases with $Re=17000$ and with no viscosity; and seven levels for $Re=17000$. Besides, we present $\mathbf{B}$ and the velocity (with vorticity) fields in the xy-plane. The velocity fields corresponded to $Re=1020$, $Re=5100$, $Re=17000$ and no viscous scenarios. We observed the formation of a bow shock with $\rho\approx\unit[3.0\times 10^{-23}]{g~cm^{-3}}$ and $p\sim\unit[1.0\times 10^{-8}]{dyn~cm^{-2}}$ and a wake with $\rho\sim\unit[10^{-24}]{g~cm^{-3}}$ and $p=\unit[1.0\times 10^{-7}]{dyn~cm^{-2}}$ at its central line. We observed that, in our simulations with five levels of refinement, the viscosity has no noticeable effects on the density profiles. Besides, we briefly discussed the simulation with six levels of refinement (left panel of Fig.~\ref{figs:8b}) and we concluded that the solutions converge for $Re=17000$.

The interaction of the wind with the planet causes the increase in $|\mathbf{B}|$ around the body when compared to its initial values: $\sqrt{B_x^2+B_y^2}$ is higher by a factor $8.5$ and $|B_z|$ remains of the same order of magnitude when compared to the initial conditions. We investigated the possible occurrence of magnetic reconnection in a case where $\nu=\unit[300]{km^2~s^{-1}}$ (right panel of Fig.~\ref{figs:3b}).

The velocity and vorticity fields of Fig.~\ref{figs:5}, as well as Fig.~\ref{figs:9}, show that $\omega_z$ slightly increases as $Re$ increases while $L$ remains approximately with the same size in the four cases.

In the case where we have and initial $B_x=0$ and using five levels of refinement, we observed the characteristics: the bow shock has $\rho=\unit[1.0\times 10^{-23}]{g~cm^{-3}}$ and $p=\unit[2.0\times 10^{-8}]{dyn~cm^{-2}}$; the wake is characterized by $\rho\sim\unit[10^{-24}]{g~cm^{-3}}$ and $p=\unit[7.0\times 10^{-8}]{dyn~cm^{-2}}$ at its inner regions. Contrary to the previous MHD case, these results present symmetry around $y=0$. Besides, $\sqrt{B_x^2+B_y^2}$ around the body is higher by a factor $4.1-4.3$ than the value calculated from the initial conditions, while $|B_z|$ is higher by a factor $1.9$.

Figure~\ref{figs:8} and Fig.~\ref{figs:9} show us that $L$ increases from $Re=1020$ to $Re=17000$, while $\omega_z$ increases between $Re=1020$ and $Re=5100$. For the scenario with no viscosity, $L=\unit[3.3\times 10^{9}]{cm}$. As in the previous MHD case, we observed higher refinement along the shocks, around the object and other regions where $\rho$ and $\mathbf{B}$ present variations, indicating us that PARAMESH remained stable in those cases.

The presence of an initial $B_x$ different from zero in the MHD simulations causes the loss of symmetry around the x-axis both in y and z-directions. We explained such a behavior as the action of the component $B_x$ on the portions of the fluid with $v_y\neq 0$ and $v_z\neq 0$, generating a dominant force in the $y$ and $z$-direction.

The absence of a bow shock in our purely hydrodynamic simulations is a characteristic observed in the interaction of the solar wind with the Moon found in \cite{spreiter:1970}. Still in \cite{spreiter:1970}, the interaction of the wind with Venus has some features in common with our MHD scenarios: the presence of a bow shock and the formation of a low-density layer between the shock and the object. In our case, this phenomenon is mainly due to the action of $|\mathbf{B}|$ around the body, while in \cite{spreiter:1970} it is caused by the ionosphere of the planet.

The influence of the viscosity on $L$ shown in \cite{grigoriadis:2010} for the hydrodynamic case is similar to the one deduced from Fig.~\ref{figs:2}: for $Re\gtrapprox50$, higher $Re$ are related to smaller $L$; in \cite{grigoriadis:2010}, the MHD case with $Re=100$ has, generally speaking, higher $L$ when compared to the scenario with $Re=40$. In our scenario with initial $B_x=0$, $L$ increases with $Re$.

We investigated the potential influence of the size and borders of the domain on the simulations. In the case used as example, we did not observe patterns which would be caused by the influence of the boundaries; however, we deduced that the size of the domain has effect on the refinement of the solutions.

From all the discussions presented here, we concluded that, under the conditions considered in this paper, our scheme generated promising results and it creates new perspectives for using the FLASH code in realistic simulations of planets interacting with stellar winds. For example, it is known that Mercury has a tenuous exosphere which undergoes strong variations between the perihelion and the aphelion, making interesting the inclusion of objects with atmospheres in future works in order to study such scenarios. We will investigate in more details the effects of higher levels of refinement on the simulations, as well as the influence of the sizes of the domain on the results. Further, we will consider scenarios with different boundary conditions and investigate how their choice affect the results.

\begin{acknowledgements}
The authors acknowledge INPE for providing the necessary computer resources. FLASH code was in part developed by the DOE NNSA-ASC OASCR Flash Center at the University of Chicago.
\bigbreak
\noindent\textbf{Funding information}
EFDE acknowledges Conselho Nacional de Desenvolvimento Cient\'{i}fico e Tecnol\'{o}gico (CNPq, PCI/INPE program, grant 300887/2017-5); OM, MOD and ODM acknowledge Financiadora de Estudos e Projetos (FINEP, under agreement 01.12.0527.00), Funda\c{c}\~{a}o de Amparo \`{a} Pesquisa do Estado de S\~{a}o Paulo (FAPESP, grant 2015/25624-2), CNPq (grants 424352/2018-4 and 307083/2017-9) and Coordena\c{c}\~{a}o de Aperfei\c{c}oamento de Pessoal de N\'{i}vel Superior (CAPES).

\end{acknowledgements}

\appendix

\section*{Appendix 1}
\label{Ap1}
\begin{figure}[H]
\centering
\includegraphics[width=1.0\linewidth]{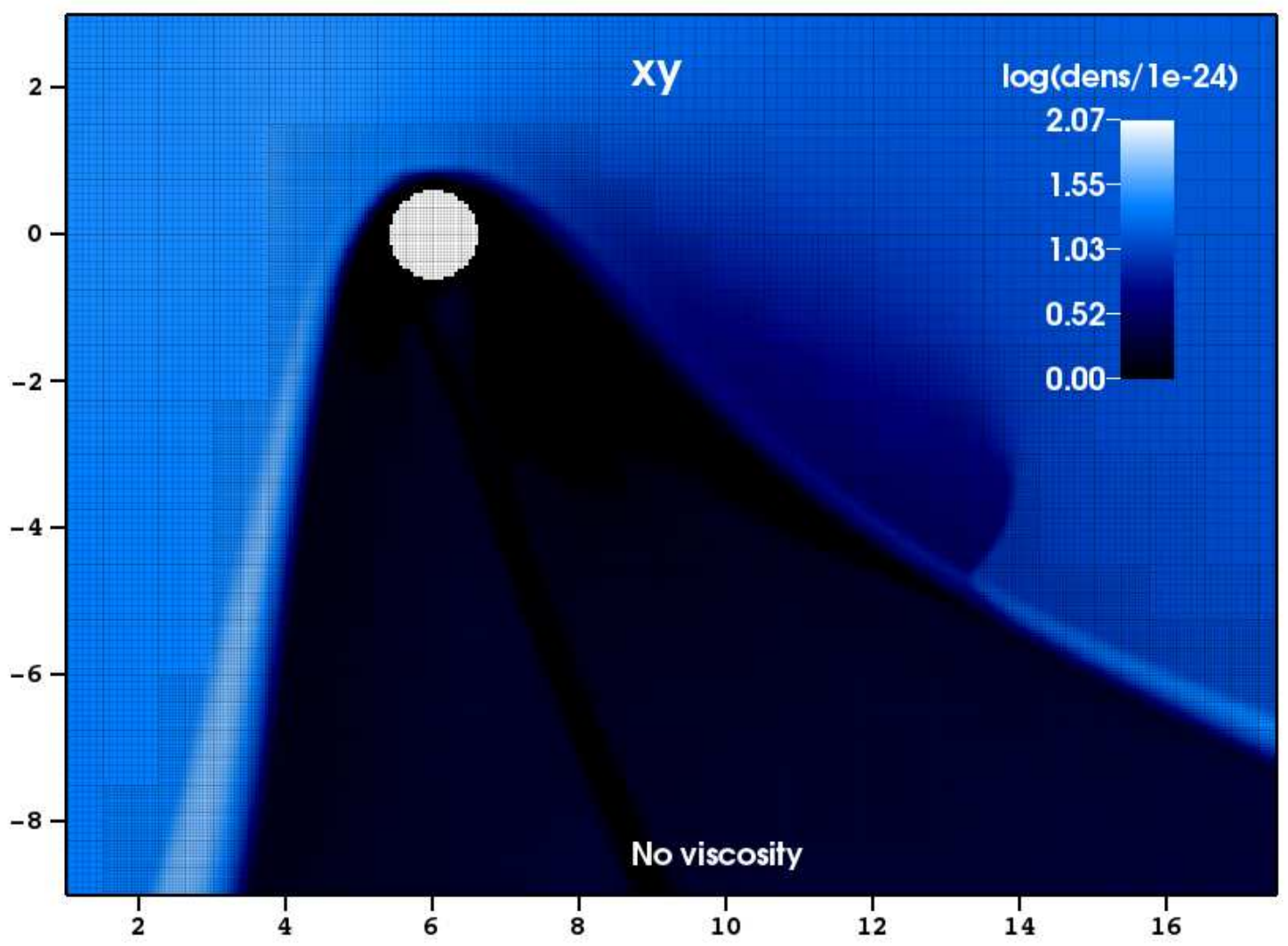}
\caption{Same as in the top left panel of Fig.~\ref{figs:3} but considering the user defined condition at the left ($x=0$), top and bottom boundaries; at the right one we use the outflow condition.}
\label{figs:10}
\end{figure}

Figure~\ref{figs:10} show the xy-plane of the simulation for a scenario similar to the one of the top panels of Fig.~\ref{figs:3} but considering the user defined condition at the left ($x=0$), top and bottom boundaries; at the right one we maintain the outflow condition.

We note that the lower region of the wake in Fig.~\ref{figs:10} is slightly wider than the one observed in the upper left panel of Fig.~\ref{figs:3}. This feature, probably, is due to the stellar wind flowing from the lower boundary, once we are considering that $v_y\neq 0$.

\section*{Appendix 2}

This Appendix yields further computational details of the simulations shown in this paper and it would be of special interest for those readers which have some familiarity with the FLASH code.

All the necessary files to the MHD simulation are placed in the folder \path{FLASH4/source/Simulation/SimulationMain/magnetoHD/StarPlanetInt}. Such files are:

\begin{itemize}
    \item \texttt{Makefile.h}: contains auxiliary instructions used to compile the particular problem being treated.
    
    \item \texttt{Config}: in this file we specify the required units and define the default runtime parameters. Particularly, we used the units \path{physics/Hydro/HydroMain/unsplit/MHD_StaggeredMesh} and \path{physics/Eos/EosMain/Gamma}.
    
    \item \texttt{flash.par}: in this file we define the initial runtime parameters such as the initial values of the physical quantities, boundary conditions, maximum level of refinement and the Riemann solver being used. See Section~\ref{numerical} for such parameters.
    
    \item \texttt{Simulation\_data.F90}: this module stores data specific to the problem being simulated.
    
    \item \texttt{Simulation\_init.F90}: this routine gets the necessary parameters and initialize other variables in the module.
    
    \item \texttt{Simulation\_initBlock.F90}: it applies the initial conditions, as well as rigid bodies and other desired particularities, to the domain of the problem. Here we insert a body in the form of a sphere of radius $R$ in the simulation by means of the algorithm:
    
    \begin{figure}[H]
    \centering
    \begin{minipage}{0.6\linewidth}
    \begin{algorithmic}
    \STATE r$=\sqrt{(x_i-x_{c})^{2}+(y_i-y_{c})^{2}+(z_i-z_{c})^{2}}$
    \STATE $\mbox{VAR(BDRY)}=-1$
    \IF {$r \leq R$}
    \STATE $\mbox{VAR(BDRY)}=1$
    \ENDIF
    \end{algorithmic}
    \end{minipage}
    \end{figure}
where $(x_i,y_i,z_i)$ and $(x_c,y_c,z_c)$ are the coordinates of the i-th cell of the domain and of the center of the sphere, respectively. The variable BDRY is defined in such a way that it has the value $+1$ in the cells inside the object; in the rest of the domain we have $\mbox{VAR(BDRY)}=-1$. Besides, inside the sphere the physical parameters have the particular values discussed in Section~\ref{numerical}.

    \item \texttt{Grid\_bcApplyToRegionSpecialized.F90}: a default version of this module is found in the folder \path{FLASH4/source/Grid}. We use it to define specific boundary conditions at the left edge of the domain, describing the stellar wind flowing toward the body, as explained in Section~\ref{numerical}.
\end{itemize}

The files \texttt{Makefile.h}, \texttt{Simulation\_data.F90} and \texttt{Simulation\_init.F90} have the standard form used in many of the supplied test problems implemented in FLASH4, which are placed in the folder \path{/FLASH4/source/Simulation/SimulationMain/}.

In order to compile and run our MHD simulation, we use the following commands:
\begin{figure}[H]
\centering
\begin{minipage}{1.0\linewidth}
\begin{algorithmic}
\STATE \verb|.\setup -auto -<n>d magnetoHD/StarPlanetInt +usm|
\STATE \verb|cd object|
\STATE \verb|make|
\STATE \verb|mpirun -np N flash4|
\end{algorithmic}
\end{minipage}
\end{figure}
\noindent
where $<$n$>$ is the number of dimensions of the simulation and N is the number of processors being used.

The files used in the pure hydrodynamic scenario are placed in \path{/FLASH4/source/Simulation/SimulationMain/StarPlanetInt}. They are similar to the ones of the MHD case, but with the following modifications:

\begin{itemize}
    \item we exclude from the files all the variables related to the magnetic field, including \texttt{killdivb}.
    
    \item in \texttt{Config} we use the unit \path{physics/Hydro/HydroMain/unsplit} instead of \path{physics/Hydro/HydroMain/unsplit/MHD_StaggeredMesh}.
\end{itemize}

To compile and run the pure hydrodynamic simulation, we use:

\begin{figure}[H]
\centering
\begin{minipage}{0.85\linewidth}
\begin{algorithmic}
\STATE \verb|.\setup -auto -<n>d StarPlanetInt +uhd|
\STATE \verb|cd object|
\STATE \verb|make|
\STATE \verb|mpirun -np N flash4|
\end{algorithmic}
\end{minipage}
\end{figure}



\end{document}